\documentclass[aps,groupedaddress,floats]{revtex4}
\usepackage{epsfig,amsmath,xspace}

\newcommand{\pct}{\%\xspace}
\newcommand{\be}{\begin{equation}}
\newcommand{\ee}{\end{equation}}

\newcommand{\adotoa}{\ensuremath{\frac{\dot a}{a}}}
\newcommand{\dlnfdlnq}[1]{\ensuremath{\frac{d \ln f_#1}{d \ln q_#1}}}
\newcommand{\lcdm}{$\Lambda$CDM\xspace}
\newcommand{\etadot}{\ensuremath{\dot \eta}}
\newcommand{\hdot}{\ensuremath{\dot h}}

\newcommand{\cps}{cosmic parameters\xspace}

\begin{document}

\bibliographystyle{revtex}

\draft

\preprint{astro-ph/yymmdddd}

\title{Probing Neutrino Properties with the Cosmic Microwave Background}
\author{Robert E. Lopez}
\email[]{r-lopez@uchicago.edu}
\thanks{this paper represents partial fulfillment of the requirements for the
Ph.D. in Physics at the University of Chicago}
\affiliation{Department of Physics \\
University of Chicago \\
Chicago, IL 60637}

\date{\today}

\begin{abstract}

Neutrinos that decay leave their imprint on the cosmic microwave background. We
calculate the CMB anisotropy for the full range of decaying neutrino parameter space,
and investigate the ability of future experiments like MAP and Planck to probe
decaying neutrino physics. We adopt two approaches: distinguishing decaying neutrino
models from fiducial \lcdm, and measuring neutrino parameters. With temperature data
alone, MAP can distinguish stable neutrino models from \lcdm if the neutrino mass
$m_h \gtrsim 2$ eV. Adding polarization data, $m_h \gtrsim 0.5$ eV is
distinguishable. Planck can distinguish $m_h \gtrsim 0.5$ eV with temperature alone,
and $m_h > 0.25$ eV with polarization. MAP without polarization can distinguish
out-of-equilibrium, early-decaying models as long as $(m_h/\rm{MeV})^2 \,
t_d/\rm{sec} \gtrsim 230$, and with polarization if $(m_h/\rm{MeV})^2 \, t_d/\rm{sec}
\gtrsim 150$. For Planck without polarization, models with $(m_h/\rm{MeV})^2 \,
t_d/\rm{sec} \gtrsim 9$ are distinguishable, and with polarization if
$(m_h/\rm{MeV})^2 \, t_d/\rm{sec} \gtrsim 6$. Models in which neutrinos decay in
equilibrium are indistinguishable from \lcdm. Late-decaying models ($10^{13} \rm{sec}
\lesssim t_d \lesssim 4 \times 10^{17} \rm{sec}$) are distinguishable from \lcdm if $m_h
\gtrsim 5$ eV for MAP and $m_h \gtrsim 2$ eV for Planck. Adding decaying neutrino
parameters to the set of \cps, we calculate the statistical uncertainty in the full
set of \cps. The ability to measure neutrino parameters depends sensitively on the
decaying neutrino model. Adding neutrino parameters degrades the sensitivity to
non-neutrino parameters; the relative amount of sensitivity degradation depends on
the decaying neutrino model, but tends to decrease with increasing experimental
sensitivity.

\end{abstract}


\maketitle

\section{Introduction}


The anisotropy in the cosmic microwave background (CMB) can be a powerful probe of
the early universe. Currently available data has already been used to place
interesting constraints on cosmic
parameters~\cite{Bunn97,Bernardis97,Lineweaver98,Hancock98,Lesgourgues98,Bartlett98,Bond98,Webster98,White98,Ratra99,Eisenstein98,Tegmark99},
and with the advent of exquisitely sensitive satellite-based experiments like
MAP~\cite{map} and Planck~\cite{planck}, it is possible to envision using the CMB to
go beyond standard parameter estimation. Many such examples have been considered:
detecting finite-temperature QED effects~\cite{Lopez98a}, constraining variations in
the fine-structure constant~\cite{Kaplinghat98}, placing limits on lepton
asymmetry~\cite{Kinney99}, and constraining Brans-Dicke
theories~\cite{Liddle98}. Another possibility is to use the CMB to probe decaying
neutrinos.

Decaying neutrinos have been considered in several cosmological contexts such as
big-bang
nucleosynthesis~\cite{Kawasaki93,Dodelson94,Dodelson94a,Madsen92,Kawasaki94},
large-scale structure formation~\cite{Bond83,Steigman85,Bond91,White95,Bharadwaj98}
and the CMB. The CMB anisotropy for models in which the neutrino decays before
recombination, $t_d \ll t_{rec} \simeq 10^{13}$ sec, have been
calculated~\cite{White95,Hannestad98,Hannestad98b,Hannestad99}. Current CMB data have
been used to study to late-decaying models, where $t_d \gtrsim 10^{13}$ sec, with the
result that masses $m_h > 100$ eV are mostly
excluded~\cite{Lopez99a,Lopez98}. Late-decaying neutrinos have also been studied in
the context of future CMB experiments~\cite{Hannestad99}. However, as pointed out in
reference~\cite{Lopez99a}, previous calculations all treated the decay radiation
perturbations as equivalent to those of the massless neutrinos. This approximation is
only valid for early-decaying scenarios. A systematic study of the CMB anisotropy in
decaying neutrino models is needed.

This work explores the use of future CMB observations, like the MAP and Planck
experiments, to probe decaying neutrino physics over a large range of neutrino
parameter space. It is organized as follows: First, we briefly discuss models of
neutrino decay in Sec.~\ref{sc:mdns}. We describe the extra steps required to
calculate the CMB spectra in Sec.~\ref{sc:cl}. In Sec.~\ref{sc:analyze}, we briefly
review cosmic parameter estimation and ruling out models in the linear regime; the
formalism used to rule out models is developed in the Appendix. We discuss how the
physics of CMB anisotropy varies as a function of neutrino mass and lifetime in
Sec~\ref{sc:regions}. This is used to break the neutrino parameter space into regions
where the physics of the CMB anisotropy is similar. We then present results:
distinguishing decaying neutrino models from standard models, and measuring cosmic
parameters in Sec.~\ref{sc:results}.

\section{Massive Neutrinos}
\label{sc:mdns}
The evidence for neutrino mass from atmospheric, solar and direct-beam neutrino
oscillation experiments is compelling\footnote{See Ref.~\cite{Raffelt99} for a review
of neutrino oscillation experimental results.}, and massive neutrinos tend to decay
unless protected by some symmetry. It is therefore interesting to consider the
cosmological signature of decaying neutrinos.

In this work we will consider neutrinos that decay non-radiatively into light decay
products. By non-radiative we mean that the decay products are electromagnetically
non-interacting. Radiative decay channels could also exist, e.g., $\nu_h \rightarrow
\nu_l \gamma$ or $\nu_h \rightarrow e^+ e^- \nu_l$. However, these models are
generally excluded by observations unless the lifetimes are extremely
long~\cite{Raffelt96,Mohapatra98}; the region of parameter space that can be probed
by the CMB is certainly excluded. There are several models with non-radiative decay
products that are motivated by particle physics. For example, familon
models~\cite{Wilczek82,Reiss82,Gelmini83} predict the following decay process: $\nu_h
\rightarrow \nu_l \phi$ where $\phi$ is a familon, a massless Nambu-Goldstone boson
associated with spontaneous breaking of a continuous, global family symmetry. In
these models, the decaying neutrino mass and mean lifetime are related at tree level
by
\begin{equation}
t_d = \frac{16\pi}{m_h} \left(\frac{F}{m_h}\right)^2
\,,
\end{equation}
where $F$ is the energy scale at which the family symmetry is broken, and it is
assumed that the neutrino $\nu_l$ is much lighter. This interaction induces a
corresponding charged-lepton decay, and experimental constraints on their branching
ratios can be used to set lower bounds on $F$. Familons corresponding to a
$\mu$-$\tau$ family symmetry are the least well constrained: the branching ratio
$B(\tau \rightarrow \mu \phi) \lesssim 3 \times 10^{-3}$~\cite{argus95} which implies
that $F \gtrsim 4 \times 10^6$ GeV for the second-third family symmetry. This leads
to the following constraint, assuming that $\nu_h = \nu_\tau$ and $\nu_l = \nu_\mu$:
\begin{equation}
\left(\frac{t_d}{\rm{sec}}\right) \, \left(\frac{m_h}{\rm{eV}}\right)^3
  > 3.0\times 10^{17} 
\,.
\end{equation}
Much of the decaying neutrino parameter space that can be probed by the CMB satisfies
this constraint.

In models where neutrinos acquire mass through the see-saw mechanism, the three-body
decay $\nu_h \rightarrow \nu_l \nu_l \bar\nu_l$ can occur~\cite{Mohapatra98},
mediated by the exchange of a $Z$-boson. However, the lifetime for this decay,
\begin{equation}
\frac{t_d}{\rm{sec}} \sim 10^{30} \left(\frac{\rm{eV}}{m_h}\right)^5
\,,
\end{equation}
is so large that the neutrino is effectively stable over the interesting region of
parameter space.

Motivated by the discussion above, we consider the following decay channel throughout
this work: $\nu_h \rightarrow \nu_l \phi$. However, alternate decay processes, like
the aforementioned $\nu_h \rightarrow \nu_l \nu_l \bar \nu_l$ do not alter our
results much; the small differences are discussed where they exist. Therefore, it is
appropriate to specify decaying neutrino models by $m_h$ and $t_d$ alone. The results
are then model-independent for most of the interesting parameter space. 


\section{Calculating the Anisotropy}
\label{sc:cl}

The anisotropy in the effective temperature of the CMB radiation, $\delta T$, is
typically described in terms of spherical harmonics,
\begin{equation}
\frac{\delta T(\theta,\phi)}{T_0} = \sum_{lm} a^T_{lm} Y_{lm}(\theta,\phi)
\,,
\end{equation}
where $\theta$ and $\phi$ describe the position on the sky, and $T_0 = 2.728$ K is
the mean background temperature of the CMB. A given theory, specified by some set of
cosmic parameters, makes predictions about the distribution of the coefficients
$a^T_{lm}$. For Gaussian theories like inflation, the coefficients are drawn from a
normal distribution, with zero mean. In this case, all of the predictions of the
theory are encoded in their variance. Therefore, the predictions of the theory can be
written in terms of the $C_l$ coefficients, defined by
\begin{equation}
C_{Tl} \equiv \left< a^T_{lm} \, a^{T\ast}_{lm} \right>
\,.
\end{equation}

In general, the temperature anisotropy does not contain all of the information in the
CMB because the CMB is polarized. The symmetric, trace-free polarization tensor
$\mathcal{P}_{ab}$ can be decomposed into two kinds of scalar modes with opposite
parities: an electric-type mode and a magnetic-type mode~\cite{Kamionkowski97}. The
polarization field can be expanded in terms of electric and magnetic type spherical
harmonics $Y^{E,B}_{lm(ab)}$, with parity $(-1)^l$ and $(-1)^{l+1}$ respectively:
\begin{equation}
\frac{\mathcal P_{ab}(\theta,\phi)}{T_0} = \sum_{lm} \left[
  a^E_{lm} Y^E_{lm(ab)}(\theta,\phi) +
  a^B_{lm} Y^B_{lm(ab)}(\theta,\phi) 
  \right]
\,.
\end{equation}
When polarization is included, the information in the CMB anisotropy can be
characterized by three additional correlation functions
\begin{align}
C_{El} &\equiv \left< a^E_{lm} \, a^{E\ast}_{lm} \right> \,, \nonumber \\
C_{Bl} &\equiv \left< a^B_{lm} \, a^{B\ast}_{lm} \right> \,, \nonumber \\
C_{Cl} &\equiv \left< a^T_{lm} \, a^{E\ast}_{lm} \right> \,.
\end{align}
Because the magnetic mode has parity opposite the temperature and electric modes, the
$T-B$ and $E-B$ correlation functions vanish~\cite{Zaldarriaga97a}. In this work we
assume that the primordial perturbations are purely scalar density perturbations,
with no tensor component. Their lack of handedness implies that scalar density
perturbations cannot generate the magnetic-type modes~\cite{Zaldarriaga98}. Therefore
$C_{Bl} = 0$ for the models we will consider. This assumption is motivated by the
fact that most inflationary models produce tensor fluctuations too small to be easily
detected, even with future satellite-based experiments~\cite{Lyth98}. In any case,
for simplicity we will ignore this possibility.

The CMB anisotropy is related to perturbations to the photon distribution function,
which is itself coupled to other particle species and gravitational metric
perturbations through particle interactions and gravity. In this work we use the
synchronous-gauge, where the coordinate and proper time of freely-falling observers
coincide; all of the metric fluctuations occur in the spatial part of the metric,
$ds^2 = a^2(\tau) \left[ -d\tau^2 + \left(\delta_{ij} + h_{ij} \right) dx^i dx^j
\right]$. The metric perturbations $h_{ij}$ can be decomposed into scalar, vector and
tensor components; we will be concerned solely with the scalar perturbations. These
can be written in terms of two scalar functions $h$ and $\eta$~\cite{Ma95}. In
Fourier space,
\begin{equation}
h_{ij} (\vec{k},\tau) = 
\left\{ \hat k_i \hat k_j \, h(\vec k,\tau) + 
  \left[ \hat k_i \hat k_j - 2 \delta_{ij} \, \eta(\vec k,\tau) \right]
\right\}
\,,
\end{equation}
where $\vec{k}$ is the Fourier mode and $\tau$ is conformal time defined in terms of
regular time $t$ and the scale factor $a$ by the relation $d\tau = d t/a$. To
calculate the CMB anisotropy we need to know the metric perturbations $h$ and $\eta$,
as well as the distribution functions for all components: decaying neutrinos, decay
radiation, photons, massless neutrinos and cold dark matter (CDM). The differences
between a standard scenario with no decaying neutrinos, and the decaying neutrino
scenarios we consider can be summarized as follows. In a decaying neutrino model:
\begin{itemize}
\item The energy densities of some of the components evolve differently from the
standard case. This affects the dynamics of the expansion of the universe through the
Friedmann equation, i.e., the Hubble parameter $\dot a / a$ is modified. This
modification is covered in Secs.~\ref{sc:friedman} and \ref{sc:endens}.
\item The Boltzmann equations that govern the evolution of the decaying neutrino and
decay radiation perturbations must be modified to include decay terms. This is
covered in Sec.~\ref{sc:pertbe}.
\end{itemize}

\subsection{Friedmann equation}
\label{sc:friedman}

The evolution of the scale factor is governed by the Friedmann equation. For the flat
universes considered here,
\begin{equation}
\left(\adotoa\right)^2 = \left(\frac{a}{a_0} \right)^2
  \frac{8\pi}{3 m_p^2} \rho(a)
\label{eq:friedman} \,,
\end{equation}
where $m_p = 1.221 \times 10^{22}$ MeV is the Plank mass and $\rho(a)$ is the total
energy density. In this work, overdots are used to denote derivatives with respect to
conformal time. The total density can be broken into components: the decaying
neutrino $\rho_h$, its decay products $\rho_{rd}$, standard radiation $\rho_{sr}$,
i.e., photons and two massless species of neutrinos, CDM + baryons $\rho_m$, and
vacuum energy density $\rho_\Lambda$. The standard components evolve simply with
scale factor: $\rho_{sr} \propto a^{-4}$, $\rho_m \propto a^{-3}$, $\rho_\Lambda
\propto a^0$. However, decays (and possible inverse decays) complicate the decaying
neutrino and decay product density evolution, which complicates the Friedmann equation
and makes it impossible to solve analytically, except in special cases.

\subsection{Energy density evolution equations} 
\label{sc:be}
\label{sc:endens}

The distribution function for the $i$-th component, $f_i(x^j,q^j,\tau)$ depends in
general on seven variables: position $x^j$, comoving momentum $q_i^j = a p_i^j$ where
$p_i$ is the proper momentum, and conformal time $\tau$; it evolves according to the
Boltzmann equation,
\begin{equation}
\frac{d f_i}{d \tau} = \frac{\partial f_i}{\partial \tau} + 
  \frac{d x^j}{d \tau} \frac{\partial f_i}{\partial x^j} + \frac{d n^j}{d \tau}
  \frac{\partial f_i}{\partial n^j} + \frac{d q}{d \tau} \frac{\partial
  f_i}{\partial q} =
a C[f_i^0]
\label{eq:be} \,,
\end{equation}
where $n^j$ is a normalized vector in the direction of the momentum, $q_i^j = q_i
n^j$, and $C[f_i^0]$ is a collision functional that describes particle
interactions. The factor of $a$ multiplying the collision functional is just
convention; it is a conversion between conformal time and real time, where collision
terms are more easily described.

To find the equations governing the evolution of the energy densities, we consider
the Boltzmann equation for the zeroth order distribution function, $f^0_i(q,\tau)$,
denoted with a superscript-$0$. By zeroth order, we mean that we are neglecting the
spatial perturbations in the distribution functions, so that the term proportional to
$\partial f^0_i / \partial x^i = 0$. The quantity $d q / d \tau$ is first order in
the metric perturbations~\cite{Ma95}, so that it too can be neglected. We also assume
that $f_i^0$ does not depend on the momentum direction ($\partial f^0_i / \partial
n^j = 0$), but allow $f_i^0$ to have arbitrary dependence on $q_i$. In this limit the
Boltzmann equation simplifies:
\begin{equation}
\frac{\partial f_i^0}{\partial \tau} = a C[f_i^0]
\label{eq:be0} \,.
\end{equation}
For the decay process $\nu_h \rightarrow \nu_l \, \phi$, the component $i$ is either
the decaying neutrino ($i\rightarrow h$) or one of the decay products ($i\rightarrow
l$, or $i\rightarrow \phi$). We can find the zeroth-order energy density from the
distribution function using the definition
\begin{equation}
\rho_i^0 = \frac{1}{a^4} \frac{g_i}{2\pi^2} \int_0^\infty \, dq \, q^2 \, 
  \epsilon \, 
  f^0_i(q) \,,
\end{equation}
where $\epsilon_i = \sqrt{q_i^2 + a^2 m_i^2}$ and $g_i$ is the number of internal
degrees of freedom for particle $i$. For the massive and massless neutrinos, $g_h =
g_l = 2$, and for the scalar decay particle $g_\phi = 1$ since it is assumed to be
spin-0 and its own antiparticle.

We next turn to the collision terms. In general, every type of interaction that the
particle experiences will contribute to the these terms. Fortunately, in the case of
decaying neutrinos, only a few interactions are important. Because the decaying
neutrino interacts with the rest of the universe via the weak interaction, it
decouples at a very high temperature of order a few MeV, just like standard, massless
neutrinos. So for temperatures of interest here (eV-scale rather than MeV-scale), the
decaying neutrino-decay radiation system is decoupled from the rest of the
universe. Therefore, the only processes that are important are decays and inverse
decays. Scatterings can be neglected in the calculation of energy densities, since
they just shuffle energy among particles~\cite{Kawasaki93}.

For the massive neutrino ($i\rightarrow h$) the collision functional can be
written~\cite{Kawasaki93},
\begin{align}
C_h[f_h^0] &= -\Gamma_D^h + \Gamma_{ID}^h 
\label{eq:cefirst} \,, \nonumber \\
\Gamma_D^h &= \frac{1}{t_d} \frac{a\,m_h}{\epsilon_h\,q_h} f_h^0(q_h) 
  \int_{1/2(\epsilon_h-q_h)}^{1/2(\epsilon_h+q_h)} 
  \,dq_1\, \left[1 + f_\phi^0(q_1)\right] \left[1 -
  f_l^0(\epsilon_h-q_1)\right] 
\,, \nonumber \\
\Gamma_{ID}^h &= \frac{1}{t_d} \frac{a\,m_h}{\epsilon_h\,q_h}
  \left[ 1 - f_h^0(q_h) \right]
 \int_{1/2(\epsilon_h-q_h)}^{1/2(\epsilon_h+q_h)} \,dq_1\, f_\phi^0(q_1)
  f_l^0(\epsilon_h-q_1) \,.
\end{align}
In this expression $\Gamma_D^h$ arises from decays, $\nu_h \rightarrow \nu_l \phi$
and $\Gamma_{ID}^h$ arises from inverse decays, $\nu_l \phi \rightarrow \nu_h$. The
integration limits follow from the kinematics of the interactions.

The collision terms for the decay products are similar to those for the decaying
neutrino. For the light neutrino ($i \rightarrow l$),
\begin{align}
C_l[f_l^0] &= -\Gamma_D^l + \Gamma_{ID}^l 
\,, \nonumber \\
\Gamma_D^l &= \frac{1}{t_d} \frac{a\,m_h}{q_l^2} \left[1 - f_l^0(q_l) \right]
 \int_{a^2m_h^2/4q_l^2}^{\infty} \, dq_\phi \, \left[1 + f_\phi^0(q_\phi) \right]
  f_h^0\left( \sqrt{ (q_\phi + q_l)^2 - a^2\,m_h^2 } \right)
\,, \nonumber \\
\Gamma_{ID}^l &= \frac{1}{t_d} \frac{a\,m_h}{q_l^2} f_l^0(q_l) 
 \int_{a^2m_h^2/4q_l^2}^{\infty} \,dq_\phi\, f_\phi^0(q_\phi)
  \left[ 1 - f_h^0\left( \sqrt{ (q_\phi + q_l)^2 - a^2\,m_h^2 } \right) \right]
\,,
\end{align}
and, for the scalar particle ($i \rightarrow \phi$),
\begin{align}
C_\phi[f_\phi^0] &= -\Gamma_D^\phi + \Gamma_{ID}^\phi
\,, \nonumber \nonumber \\
\Gamma_D^\phi &= 2 \frac{1}{t_d} \frac{a\,m_h}{q_\phi^2} 
  \left[1 + f_\phi^0(q_l) \right]
 \int_{a^2m_h^2/4q_\phi^2}^{\infty} \, dq_l \, \left[1 + f_\phi^0(q_\phi) \right]
  f_h^0\left( \sqrt{ (q_l + q_\phi)^2 - a^2\,m_h^2 } \right)
\,, \nonumber \\
\Gamma_{ID}^l &= 2 \frac{1}{t_d} \frac{a\,m_h}{q_\phi^2} f_\phi^0(q_\phi) 
  \int_{a^2m_h^2/4q_\phi^2}^{\infty} \,dq_l \, f_l^0(q_l)
  \left[ 1 - f_h^0( \sqrt{ (q_\phi + q_l)^2 - a^2\,m_h^2 } ) \right]
\label{eq:celast} \,.
\end{align}

The Boltzmann equation for each type of particle, Eq.~\ref{eq:be0}, their collision
term equations, Eqns.~\ref{eq:cefirst}--\ref{eq:celast}, and the Freidman equation,
Eq.~\ref{eq:friedman}, determine the dynamics of the expansion of the universe. They
form a closed set of integro-differential equations for the evolution of the scale
factor, and require numerical methods for their solution. In particular, the
collision term integrals are complicated functions of momentum. However, for certain
special cases these equations simplify, and for other cases we can estimate the
late-time densities without having to solve the equations at all.

\subsubsection{Out-of-equilibrium decays}

Neutrino decays become important when the age of the universe is near the neutrino
mean lifetime. If $T(t_d) \ll m_h/3$, where $T(t_d)$ is the temperature of the
universe at time $t_d$ after the big-bang, the neutrino decays non-relativistically,
so that when the neutrino starts to decay, the thermal energy of the decay products
cannot overcome the rest mass energy of the decaying neutrinos. This suppresses
inverse decays relative to decays and causes the decays to occur out of
equilibrium. We will use the terms out-of-equilibrium decays and non-relativistic
decays interchangeably. Thus, the neutrino decays away when $t \sim t_d$. These
decays can generate a large amount of decay radiation, depending on the initial
abundance of the decaying neutrino and how non-relativistic the neutrino is at decay.

Since neutrinos decouple from the rest of the universe at a very high temperature
determined by their weak interactions, all of the neutrinos, including the massive,
decaying neutrino, are ultra-relativistic at decoupling (we will not consider
MeV-scale decaying neutrinos). Thus, their abundances are large, of order the photon
abundance. The decaying neutrino, if still present, becomes non-relativistic when $T
\lesssim m_h/3$. Then its energy density scales as matter, as $a^{-3}$ instead
of as radiation, which scales as $a^{-4}$. Its energy density, and consequently the
energy density of its decay products, becomes relatively more important the longer
the decaying neutrino is still around and non-relativistic.

For out-of-equilibrium decays, simplified evolution equations for the decaying
neutrino and decay radiation densities can be found. The collision term for the
decaying neutrino simplifies, because in this limit we can neglect $f_l^0$ and
$f_\phi^0$:
\begin{equation}
C_h[f_h^0] \simeq \frac{q}{t_d} \, \frac{m_h}{\epsilon_h} \, f_h^0
\,.
\end{equation}
The Boltzmann equation can then be converted into a differential equation for
$\rho^0_h$, by multiplying each term by $p_h^2 E_h$ and integrating out $p_h$. We
find that
\begin{equation}
\frac{\partial \rho_h^0}{\partial \tau} + 3 \adotoa 
  \left(\rho_h^0 + P_h^0\right) = 
  - \frac{a m_h}{t_d} \, n_h^0
\label{eq:dh} \,,
\end{equation}
where $P_h^0$ is the pressure and $n_h^0$ is the number density of the decaying
neutrinos, given by the definitions,
\begin{align}
n^0_i &= 
  \frac{1}{a^3} \frac{g_i}{2\pi^2} 
  \int_0^\infty \, dq \, q^2 \, f^0_i(q) 
\,, \nonumber \\ 
P^0_i &= 
  \frac{1}{a^4} \frac{g_i}{2\pi^2} 
  \int_0^\infty \, dq \, \frac{q^4}{3 \epsilon} \, f^0_i(q)
\,.
\end{align}
A couple of comments about the evolution equation for $\rho^0_h$ are in order. Note
the presence of a pressure term $P^0_h$ on the left hand side. In the limit of
completely non-relativistic decays, this term is zero, but otherwise this term can be
a significant correction. If we neglect the pressure term, then the $3 (\dot a / a)
\rho_h$ term represents the fact that matter density varies as $a^{-3}$ in the
absence of decays. A similar comment applies to the product $m_h \, n^0_h$ on the
left hand side. For completely non-relativistic decays, all of the decaying neutrino
energy is rest mass energy so that $m_h \, n^0_h = \rho^0_h$, but otherwise the two
quantities are not equal.

Given the decaying neutrino energy density, the decay product energy density
$\rho_{rd} = \rho_l + \rho_\phi$ can be obtained from the first law of
thermodynamics~\cite{Kolb90}:
\begin{equation}
\frac{d (a^3 \rho_{rd})}{d \tau} = 
  - P_{rd} \frac{d a^3}{d \tau} - \frac{d (a^3 \rho_h) }{d \tau}
\,.
\end{equation}
Since the decay radiation is massless, $P_{rd} = 1/3 \rho_{rd}$, and we find that
\begin{equation}
\frac{d \rho_{rd}}{d \tau} + 4 \adotoa \rho_{rd} = 
  - \left( \frac{d \rho_h}{d \tau} + 3 \adotoa \rho_h \right) = - \frac{a}{t_d}
  \rho_h
\label{eq:drd} \,,
\end{equation}
where the second equality holds for fully non-relativistic decays. In the absence of
decays, this equation implies that the decay radiation density scales as $a^{-4}$, as
expected for massless particles. Finally, we can obtain a simpler equation, which
will be useful later, for the evolution of the decay radiation density. Let $r_{rd} =
\rho^0_{rd} / \rho^0_{1\nu}$, where $\rho^0_{1\nu}$ is the density in a single
species of standard, massless neutrinos. Then Eq.~\ref{eq:drd}, and the fact that
$\rho^0_{1\nu} \propto a^{-4}$, implies that
\begin{equation}
\frac{dr_{rd}}{d\tau} = 
\frac{m_h\,n_h^0}{\rho_{1\nu}}\,\frac{a}{t_d}
\,.
\end{equation}

\begin{figure}
\centerline{\epsfig{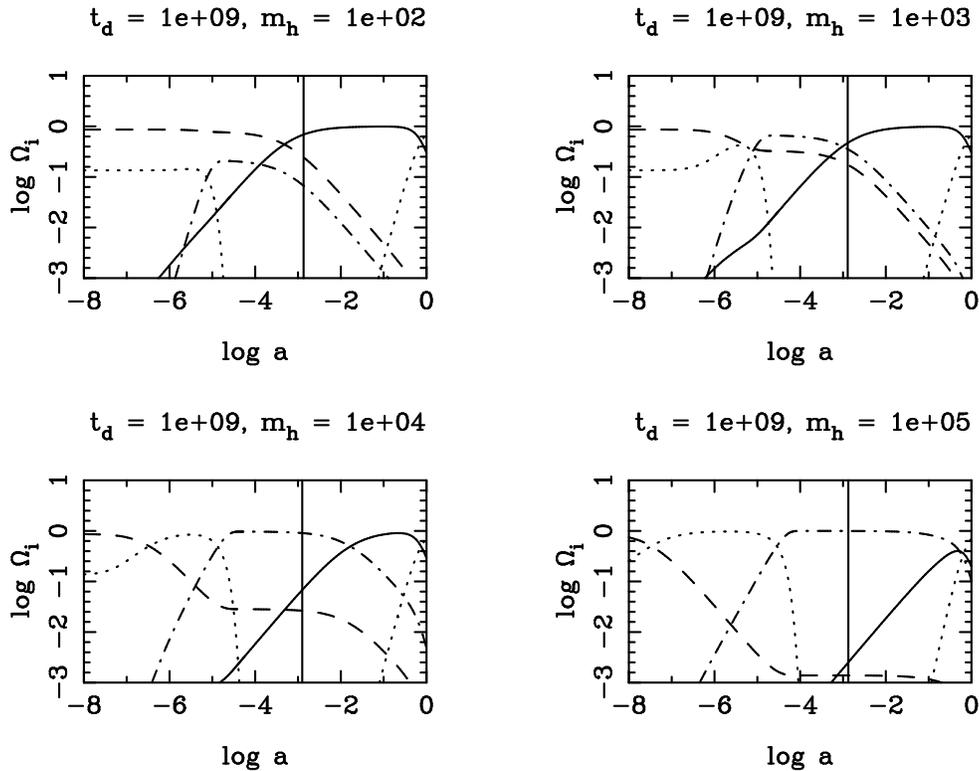}}
\caption{The evolution of the energy densities, relative to the critical density, of
the various components of the universe, in early-decaying scenarios. The notation is
as follows: solid line = CDM+baryons, long-dashed line = standard radiation (photons
+ 2 massless neutrinos), dotted line = decaying neutrino, dot-dashed line = decay
radiation. For all values of the scale factor, $\sum_i \Omega_i = 1$. The background
cosmological model has a cosmological constant $\Omega_\Lambda = 0.7$ today. The
vertical line represents the epoch of recombination. The models shown here all have
$t_d = 10^9$ sec; $m_h$ varies from $10^2$ to $10^5$ eV. For $m_h = 10^2$ eV, the
decay is barely non-relativistic: $\alpha = 1.1$. The decay radiation density never
matches the density in standard radiation. For the higher-mass scenarios, the decays
are out-of-equilibrium and the decay radiation dominates the standard radiation for
all times after decay. Another feature to be noted is the relative importance of
components at recombination; this determines the amount of the early-ISW effect. For
$m_h = 10^4$, $10^5$ eV, the universe is radiation dominated at last scattering,
creating a large early-ISW effect.}
\label{fg:omegaTCDM}
\end{figure}

\begin{figure}
\centerline{\epsfig{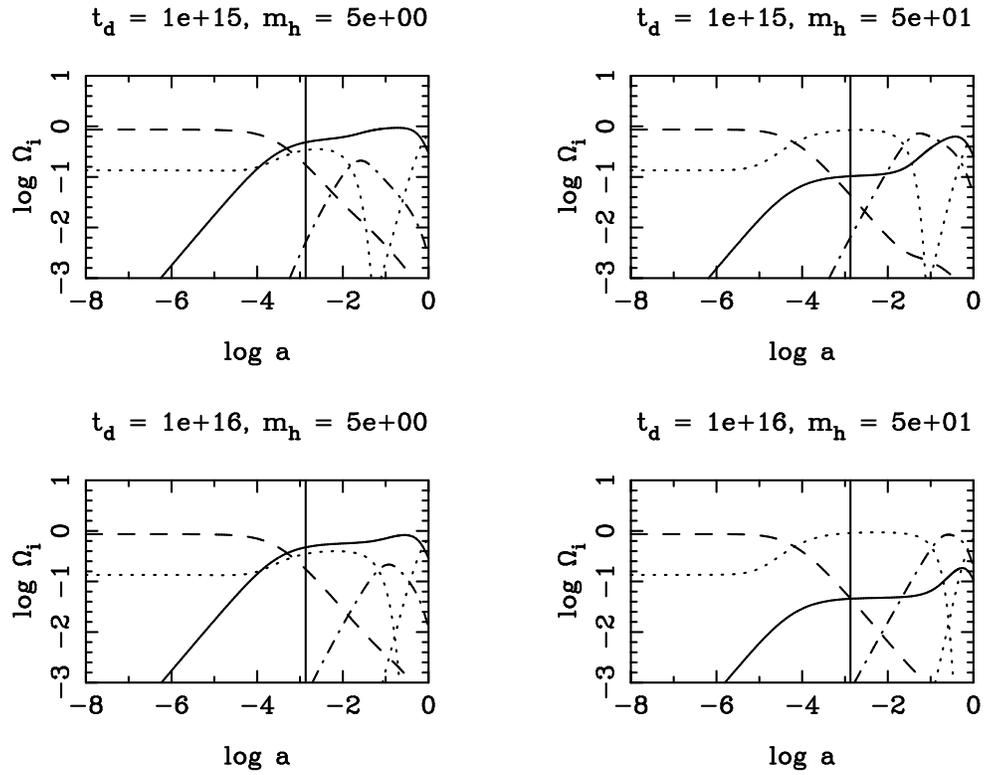}}
\caption{Same as Fig.~\ref{fg:omegaTCDM}, but for late-decaying neutrinos. In neither
$m_h = 5$ eV scenario does either the massive neutrino or its decay products ever
dominate the energy density. In both scenarios with $m_h = 50$ eV, the universe is
dominated by the massive neutrino at recombination, and by the decay radiation at
decay.}
\label{fg:omegaLate}
\end{figure}

To find the energy densities of the decaying neutrino and its decay radiation for
out-of-equilibrium decays, we numerically solve Eqns.~\ref{eq:dh}~and~\ref{eq:drd},
together with the Friedmann equation, Eq.~\ref{eq:friedman}~\footnote{The Boltzmann
code used in this calculation was written by Robert Scherrer.}. Results for several
decaying neutrino models are shown in
Fig's~\ref{fg:omegaTCDM}~and~\ref{fg:omegaLate}. There we plot the energy densities,
scaled by the critical density, for all of the components: standard radiation, CDM,
vacuum energy density, decaying neutrino and decay radiation. The first figure shows
a succession of masses with lifetimes fixed at $10^9$ sec. These are models where the
neutrino decays before last scattering, $t_{rec} \sim 10^{13}$ sec. It is easy to see
that the decay radiation becomes more important as the mass increases, in keeping
with Eq.~\ref{eq:nrd}. If the neutrino is massive enough, then it can cause an early
phase of matter domination before it decays and its decay radiation dominates. The
second figure shows some models where the neutrino decays after last scattering.

\begin{figure}[!ht]
\centerline{\epsfig{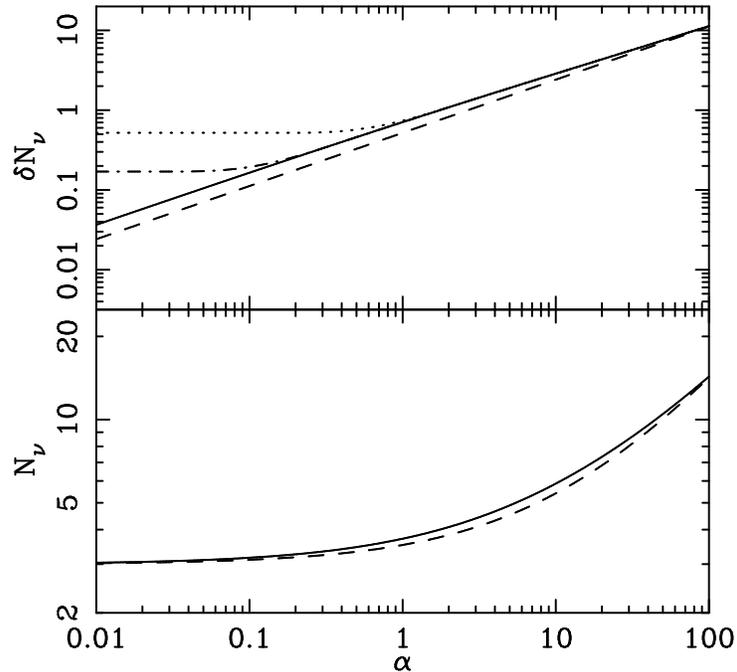}}
\caption{Late-time asymptotic behavior of extra radiation energy density expressed in 
units of species of massless neutrinos, as a function of relativity parameter
$\alpha$. Plotted are the non-equilibrium limit $\delta N_\nu = 0.52 \alpha^{3/2}$,
valid for $\alpha \gg 1$, as well as data for 2-body decays ($\nu_h \rightarrow \nu_l 
\,\phi$) and 3-body decays ($\nu_h \rightarrow \nu_l\,\nu_l\,\nu_l$). In
both the 2 and 3-body decay scenarios we have assumed an initial thermal abundance of
heavy and light neutrinos at the standard neutrino temperature $T_\nu = (4/11)^{1/3}
T_\gamma$.}
\label{fg:dn_a}
\end{figure}

By determining how non-relativistic the neutrino is when it decays, it is possible to
obtain an estimate of the energy density in decay radiation, without resorting to the
full Boltzmann equations. To do this we define a relativity parameter $\alpha$,
proportional to the square of the decaying neutrino's mass divided by its thermal
energy at the time of decay, and with the property that $\alpha
\simeq 1$ at the border between relativistic and non-relativistic decays. For
non-relativistic decays, $\alpha \gg 1$ and $\rho_{rd} / \rho_{1\nu}$ is large; for
ultra-relativistic decays, $\alpha \ll 1$ and $\rho_{rd} / \rho_{1\nu} \simeq
1$. Consider a scenario with $\alpha \simeq 1$. Here, the universe is never dominated
by the massive neutrino. For a radiation dominated universe at decay, the Friedmann
equation gives the relation between decay time and temperature~\cite{Kolb90}, $t_d
\simeq 0.3 g_\ast^{-1/2} m_p/T_d^2$, where $g_\ast \simeq 3.36$ is the effective
number of relativistic degrees of freedom. Since the temperature at decay $T_d \simeq
m_h/3$, the neutrino parameters enter in the combination $m_h^2 t_d$, which implies
that
\begin{equation}
 \alpha = 0.11 \,
  \left(\frac{m_h}{\rm{MeV}}\right)^2 \left(\frac{t_d}{\rm{sec}}\right)
\,. \label{eq:relparam}
\end{equation}
Because matter density decreases as one power of the scale factor relative to
radiation density, we can estimate the energy density in decay radiation in units of
standard massless neutrinos, $N_{rd}$, as follows: $N_{rd} \simeq a_d/a_{nr}$, where
$a_{nr}$ is the scale factor when the neutrino becomes non-relativistic and $a_{d}$
is the scale factor at decay. Here we assume that the decay instantaneously
transforms the density in decaying neutrinos to the decay radiation. If the universe
is dominated by the decaying neutrino at decay, then the Friedmann equation can be
used to obtain $a_{d}$. The result is that~\cite{Hannestad99}
\begin{equation}
N_{rd} \simeq 0.52 \, \alpha^{2/3}
\label{eq:nrd} \,,
\end{equation}
valid for $\alpha \gg 1$. The numerical coefficients in Eqns.~\ref{eq:relparam} and
\ref{eq:nrd}, but not the overall dependence, have been fudged by a small amount so
that the the formula for $N_{rd}$ agrees well with numerical results. The bottom pane
of Fig.~\ref{fg:dn_a} shows this estimate versus numerical results for the total
radiation density $N_\nu = N_{rd} + 2$, as a function of $\alpha$. The $2$ represents
the two species of massless neutrinos. As the figure shows, the agreement is good as
a rough estimate.

\subsubsection{Equilibrium decays}

If $T(t_d) \gg m_h/3$, neutrino decays become important while the neutrino is still
ultra-relativistic. In this case both decays and inverse decays occur, and the
collision terms do not simplify. It is, however, possible to obtain an estimate for
the energy density in decay products long after the neutrino has decayed away. This
estimate relies on the fact that when $t \gtrsim t_d$, the decay and inverse decay
processes are sufficiently fast relative to the expansion rate to establish chemical
equilibrium between the decaying neutrino and its decay
products~\cite{Madsen92}. Then the distribution functions are approximately thermal
in form with pseudo-temperature $T'$ not necessarily equal to the temperature of the
universe. Therefore, we have
\begin{equation}
f_i^0 = \frac{1}{e^{(E_i - \mu_i)/T'}\pm 1}
\,,
\end{equation}
for $i=h$, $l$, or $\phi$, with pseudo-chemical potentials related by
\begin{equation}
\mu_h = \mu_l + \mu_\phi
\label{eq:psmu} \,.
\end{equation}
It is easy to show that the Boltzmann equations imply that the following relations
hold generally~\cite{Kawasaki93}:
\begin{align}
\frac{d}{d\tau} \left(a^3 \left(n_h + n_l \right) \right) &= 0 
\label{eq:psn1} \,, \\
\frac{d}{d\tau} \left(a^3 \left(n_h + n_\phi \right) \right) &= 0
\label{eq:psn2} \,,
\end{align}
and that for $T \gg m_h$, the following holds:
\begin{equation}
\frac{d}{d\tau} \left(a^4 \left(\rho_h + \rho_l + \rho_\phi \right) \right) 
= 0
\label{eq:psrho} \,.
\end{equation}
This equation implies that the total comoving energy density in the decaying
neutrino-decay radiation system is unchanged by the decay/inverse decay processes,
for $T(t_d) \gg m_h$.

We numerically solve Eqns.~\ref{eq:psmu}, \ref{eq:psn1}, \ref{eq:psn2}, and
\ref{eq:psrho} for $\mu_h$, $\mu_l$, $\mu_\phi$, and $T'$. For initial conditions we
assume a thermal initial abundance of heavy and light neutrinos, and no initial
scalar particles. We find that
\begin{equation}
\mu_h = 0.092 \, T_\nu \,, \,\,
\mu_l = 0.581 \, T_\nu \,,  \,\,
\mu_\phi = -0.489 \, T_\nu \,, \,\,
T' = 0.884 \, T_\nu
\,,
\end{equation}
where $T_\nu = (4/11)^{1/3} T$ is the standard neutrino temperature. This solution if
valid for $t \gtrsim t_d$ and $m_h \lesssim T(t)$. When the universe cools enough so
that $T \lesssim m_h/3$, the inverse decays become suppressed and decays
predominate. The rest mass of the heavy neutrinos is starting to become important,
increasing their total energy relative to the massless case. As they decay away, the
energy density in heavy neutrinos is then transferred to the decay radiation, raising
its temperature. We can calculate the amount of heating by using the fact that the
entropy of the heavy neutrino-decay radiation system is conserved. We find that the
neutrino decays raise the decay radiation temperature by 14.7\pct. Our final result,
the energy density in decay radiation, can be expressed in units of standard,
massless neutrino energy density: $N_{rd} = 2.17$. In a scenario with no decaying
neutrinos, this number would be 2.0, so $\delta N_\nu = 0.17$, where $\delta N_\nu$
is the change in radiation density. A similar procedure could also be repeated for
the case of three body decays: $\nu_h \rightarrow 3\,\nu_l$. In this case, $\delta
N_\nu = 0.52$ for $\alpha
\ll 1$, and the large-$\alpha$ behavior of the radiation density is identical to the
two-body case.

Results for the decay radiation energy density, for both equilibrium and
out-of-equilibrium decays, are shown in Fig.~\ref{fg:dn_a}. To summarize, for
$T(t_d) \ll m_h/3$, decays occur in equilibrium, with decays and inverse decays
important for $t \gtrsim t_d$. For $t_d < t < t(T=m_h/3)$ the energy density in
radiation is repartitioned, but the total value is the same as if the neutrino did
not decay. For $t \gtrsim t(T=m_h/3)$, the heavy neutrino decays away and increases
the total radiation density by 2.3\pct.

\subsection{Perturbation Boltzmann equations}
\label{sc:pertbe}

Following reference~\cite{Ma95}, we would like to derive a hierarchy of Boltzmann
equations describing the evolution of perturbations to the decaying neutrinos and the
decay radiation. The $i$-th distribution function can be written as the product of an
unperturbed, thermal function times a perturbation, as follows:
\begin{equation}
f_i(x^j,q,n^j,\tau) = f_i^0(q,\tau) 
  \left[ 1 \pm \Psi_i(x^j,q,n^j,\tau ) \right]
\,.
\end{equation}
The Boltzmann equation, Eq.~\ref{eq:be}, then yields an equation for the evolution
of the perturbation. Upon taking the Fourier transform,
\begin{equation}
\frac{\partial \Psi_i}{\partial \tau} 
  + i \frac{q_i}{\epsilon_i} \left(\vec k \cdot \vec n\right) \Psi_i + \dlnfdlnq{i}
  \left[ \dot \eta - \frac{1}{2}(\dot h+\dot \eta)(\vec k \cdot \vec n)^2 \right] 
  = 
  \frac{1} {f^0_i} \left(\frac{\partial f}{\partial \tau}\right)_C
\,. \label{eq:pertbe}
\end{equation}

Since the decay radiation is effectively massless, $\epsilon_{rd} = q_{rd}$, and we
can integrate the momentum dependence out of the Boltzmann equation. We define a
momentum-independent perturbation $F_{rd}$ as in reference~\cite{Ma95}, scaling it by
the decay related factor $r_{rd}$ for convenience:
\begin{equation}
F_{rd}(\vec{k}, \hat{n}, \tau) \equiv
\frac{\int dq\, q^3f^0_{rd}(q, \tau)\Psi_{rd}(\vec{k}, q, \hat{n}, \tau)}
{\int dq\, q^3 f^0_{rd}(q, \tau)}\;r_{rd}
\,. \label{frd} 
\end{equation}
Unfortunately, the complicated form for the collision terms in the Boltzmann equation
makes it difficult to derive simple equations in the general case. {\em For the rest
of this section, we will specialize to the case of out-of-equilibrium decays}, where
these terms simplify. Then the Boltzmann equation governing $F_{rd}$ can be shown to
be~\cite{Lopez99a}
\begin{equation} 
\dot{F}_{rd}+\imath k\mu F_{rd}+4\left(\frac{\dot{h}}{6}+
\frac{\hdot+6\etadot}{3}
P_2(\mu)\right)r_{rd} = \dot{r}_{rd} N_0
\,, \label{eq:frddot}
\end{equation}
where
\begin{equation}
N_0(k, \tau) =     
\frac{\int dq_h\, q_h^2f^0_{h}(q_h, \tau)\Psi_{h}(k, q_h, \tau)
\left[1-\frac{8}{3}\left(\frac{q_h}{am_h}\right)^2\right]}
{\int dq_h\, q_h^2 f^0_{h}(q_h, \tau)}
\,, \label{eq:N_0} 
\end{equation}
$\mu=\hat{k}\cdot\hat{n}$ and $P_n(\mu)$ are the Legendre polynomials of order
$n$. In Eqns.~\ref{eq:frddot} and~\ref{eq:N_0}, only terms up to ${\cal
O}(q^2_h/a^2m^2_h)$ have been kept. Similar equations for the evolution of
perturbations in the decay radiation can be found in references~\cite{Bond91},
\cite{Bharadwaj98} and ~\cite{Lopez99a}.


The dependence of $F_{rd}$ on $\mu$ can be eliminated by expressing it as a series of
Legendre polynomials, $F_{rd} = \sum_l F_{rd,l} \,P_l$, leading to the following
Boltzmann hierarchy for the decaying neutrino perturbations, valid for
out-of-equilibrium decays:
\begin{align}
\dot \delta_{rd} &= -\frac{2}{3} \left(\dot h + 2\theta_{rd}\right)
  - \frac{\dot r_{rd}}{r_{rd}} \left( \delta_{rd} - \delta_h \right)
\,, \nonumber \\
\dot \theta_{rd} &= k^2 \left( \frac{1}{4} \delta_{rd} - \sigma_{rd} \right)
  - \frac{\dot r_{rd}}{r_{rd}}  \theta_{rd}
\,, \nonumber \\
\dot \sigma_{rd} &= 
  \frac{2}{15} \left( 2\theta_{rd} + \dot h + 6 \dot \eta \right) - 
  \frac{\dot r_{rd}}{r_{rd}}  \sigma_{rd}
\,, \nonumber \\
\dot F_{rd,l} &=
  \frac{k}{2l+1} \left[ l F_{rd,l-1} - (l+1) F_{rd,l+1} \right]
   - \frac{\dot r_{rd}}{r_{rd}}  F_{rd,l} \,, \;\; l\ge 3
\label{eq:drbe}
\end{align}
where $\delta_{rd}=F_{rd,0}/r_{rd}$, $\theta_{rd}=3kF_{rd,1}/4r_{rd}$ and
$\sigma_{rd}=F_{rd,2}/2r_{rd}$. This set of equations is identical to the Boltzmann
hierarchy for standard massless neutrinos, with the addition of the terms
proportional to $\dot r_{rd}/ r_{rd} \propto 1 / t_d$. The extra decay term can have
a large effect on the decay radiation perturbations when $t \sim t_d$, but for late
times the perturbations approach the values they would have attained in its absence.
To calculate the decay radiation perturbations, we added a separate Boltzmann
hierarchy, described by Eq.~\ref{eq:drbe}. In our numerical scheme, this hierarchy
must be terminated at some value of multipole moment $l_{end}$. We terminate the
hierarchy by adding the extra equation for $F_{rd,l_{end}+1}$,
\begin{equation}
F_{rd,l_{end}+1} = \frac{2 l_{end} + 1}{k\tau} F_{rd,l_{end}} - 
  F_{rd,l_{end}-1}
\,,
\end{equation}
the method used for the massless neutrino hierarchy in CMBFAST~\cite{Seljak96}.

Because the decaying neutrinos are massive, $\epsilon_h \neq q_h$, and it is
impossible to integrate the momentum dependence from their perturbations. After
expanding the decaying neutrino perturbation in terms of Legendre polynomials,
$\Psi_h = \sum_l \Psi_{h,l} P_l$, the Boltzmann equation becomes
\begin{align}
\dot \Psi_{h,0} &=
  -\frac{q_h k}{\epsilon_h} \Psi_{h,1} + \frac{1}{6} \dot h \dlnfdlnq{h} - 
  \frac{a m_h}{t_d \epsilon_h} \Psi_{h,0}
\,, \nonumber \\
\dot \Psi_{h,1} &= 
  \frac{q_h k}{3 \epsilon_h} \left( \Psi_{h,0} - 2 \Psi_{h,2} \right) - \frac{a
  m_h}{t_d \epsilon_h} \Psi_{h,1}
\,, \nonumber \\
\dot \Psi_{h,2} &= 
  \frac{q_h k}{5 \epsilon_h} \left( 2\Psi_{h,1} - 3\Psi_{h,3} \right) -
  \left(\frac{1}{15} \dot h
  + \frac{2}{5}\dot \eta \right) \dlnfdlnq{h}
  - \frac{a m_h}{t_d \epsilon_h} \Psi_{h,2}
\,, \nonumber \\
\dot \Psi_{h,l} &= 
  \frac{q_h k}{\epsilon+h \left(2l+1\right)} \left[l \Psi_{h,l-1} - (l+1)
  \Psi_{h,l+1} \right] 
  - \frac{a m_h}{t_d \epsilon_h} \Psi_{h,l}
\,, \;\; l \ge 3
\,.
\end{align}
This set of equations differs from the evolution equations for massive, non-decaying
neutrinos only through the presence of the term proportional to $1/t_d$. The decay
term is easily interpreted. For non-relativistic neutrinos, the $m_h$ in the
numerator cancels the $\epsilon_h$ in the denominator; the result is just the
differential equation for exponential decay, in conformal time. If the neutrinos are
not completely non-relativistic, then their velocities become important, and there is
a time dilation factor associated with transforming between the neutrino rest frame
and the thermal frame. In this case, $m_h / \epsilon_h$ becomes the special
relativistic gamma factor for this transformation.

It should be noted that, for out-of-equilibrium decays, the perturbation evolution
equations are independent of the details of the decay radiation, except for the fact
that is must be light and weakly-interacting. The energy density equations,
Eq.~\ref{eq:drd} and
\ref{eq:dh}, are also independent of the details. Therefore, the CMB anisotropy
becomes a function of $m_h$ and $t_d$, independent of the decay channel. In fact,
the calculations can be generalized to encompass generic decaying particles. The main
difference in the generic scenario will be due to the initial abundance of the
decaying particle which will depend on its interactions. However, a generic decaying
particle will produce a CMB spectrum very similar to a decaying neutrino with the
same lifetime, provided that the densities of decay radiation are the same. Finally,
note that this simplification is valid for out-of-equilibrium decays only.


\section{Analyzing the Data}
\label{sc:analyze}

This section briefly reviews estimating cosmic parameter uncertainties (``error
forecasting''), and using data to rule out decaying neutrino parameter space. For
further discussion of error forecasting in parameter estimation, see e.g.,
Refs.~\cite{Jungman96,Dodelson96,Zaldarriaga97,Zaldarriaga97a,Bond97,Tegmark97}. The
formalism used to determine which decaying neutrino models can be ruled out is
discussed in the Appendix.

\subsection{Measuring $m_h$ and $t_d$}
\label{sc:fim}

A given theory, specified by a set of cosmological parameters $\{\lambda_i\}$
($i=1\ldots N$, with $N$ the number of \cps considered) makes predictions about the
multipole amplitudes, the $C_l$'s. The results of a CMB experiment are estimates of
the $C_l$'s, with some experimental uncertainties. Of course, we cannot know in
advance the values of $C_l$'s that a given experiment will measure; however, by
knowing what we expect for the uncertainties, we can estimate how large the
uncertainties in the parameters should be.

For an experiment with data out to some maximum $l=l_{max}$, we can define a goodness
of fit statistic that is a function of $\{\lambda_i\}$:
\begin{equation}
\chi^2(\{\lambda_i\}) = \sum_{l=2}^{l_{max}} \sum_{X,Y=T,E,C}\, 
  \left[ C_{Xl}^{theory}(\{\lambda_i\}) - C_{Xl}^{data} \right]
  V^{-1}_{XYl}
  \left[ C_{Yl}^{theory}(\{\lambda_i\}) - C_{Yl}^{data} \right]
\label{eq:chi2}
\,,
\end{equation}
where $C_{Xl}^{theory}$ is the theoretical spectrum for \cps $\{\lambda_i\}$,
$C_{Xl}^{data}$ is the measured spectrum and $V_{XYl}$ is the covariance matrix
between estimators of the different spectra. For a cosmic variance limited experiment
with data to some maximum $l=l_{max}$, the diagonal components of $V_{XYl}$ are given
by~\cite{Zaldarriaga97a}
\begin{align}
V_{TTl} &= \frac{2}{2l+1} C_{Tl}^2 \,, \nonumber \\
V_{EEl} &= \frac{2}{2l+1} C_{El}^2 \,, \nonumber \\
V_{CCl} &= \frac{2}{2l+1} \left( C_{Cl}^2 + C_{Tl} C_{El} \right) 
\label{eq:covdiag}
\,,
\end{align}
and the non-zero off-diagonal components are given by
\begin{align}
V_{TEl} &= \frac{2}{2l+1} C_{Cl}^2 \,, \nonumber \\
V_{TCl} &= \frac{2}{2l+1} C_{Tl} C_{Cl} \,, \nonumber \\
V_{ECl} &= \frac{2}{2l+1} C_{El} C_{Cl} \,,
\label{eq:covoffdiag}
\end{align}
for $l \le l_{max}$.

The measured \cps, $\{\lambda_i^\prime\}$, are determined by minimizing
$\chi^2(\{\lambda_i\})$:
\begin{equation}
\frac{\partial \chi^2_{min}}{\partial \lambda_j}(\{\lambda'_i\}) = 0
\,,
\end{equation}
for $j=1 \ldots N$. If we assume that the measured \cps are close to their actual
values, denoted $\{\tilde \lambda_i\}$, then $\chi^2$ can be expanded about its
minimum as follows:
\begin{equation}
\chi^2\left(\left\{\lambda_i\right\}\right) \simeq
 \chi^2 (\left\{\tilde \lambda_i\right\} )
+ \sum_{ij} \left(\lambda_i - \tilde \lambda_i\right) \alpha_{ij} 
\left(\lambda_j - \tilde \lambda_j\right)
\,,
\end{equation}
where $\alpha_{jk}$ is the Fisher matrix,
\begin{equation}
\alpha_{ij} = \sum_l \sum_{XY} 
  \frac{\partial C_{Xl}^{theory}}{\partial \lambda_i}
  V^{-1}_{XYl} 
  \frac{\partial C_{Yl}^{theory}}{\partial \lambda_j} \,.
\label{eq:fm}
\end{equation}
The Fisher matrix determines how rapidly $\chi^2$ increases as the \cps are varied
away from their true values. Under certain reasonable assumptions~\cite{Press90}, the
uncertainties on the parameters are determined by this matrix. If we allow all \cps
to vary simultaneously, then 
\begin{equation}
\delta \lambda_i = \sqrt{\left(\alpha^{-1}\right)_{ii}}
\label{eq:delcp} \,.
\end{equation}

The formalism above assumes data for both temperature and polarization. If only
temperature data is obtained, then the covariance matrix $V_{XYl}$ becomes a number:
\begin{equation}
V_{XYl} = \frac{2C_{Tl}^2}{2l+1} \, \delta_{XT} \, \delta_{YT}
\,,
\end{equation}
where $\delta$ is the discrete delta function.

To calculate the uncertainties in the parameters, we will assume some decaying
neutrino scenario. The set of \cps will include neutrino parameters, like $m_h$ and
$t_d$. The uncertainties will then depend on the model we assume and the parameters
we allow to vary.

\subsection{Ruling out models}
\label{sc:dist}

It could also be the case that no theoretical model can specify the data. For
instance, in a decaying neutrino scenario, the data could be analyzed without
considering neutrino parameters. In general, two things will then happen. 1) The
best-fit parameters will be systematically offset from the true parameters. 2) No
theoretical model will fit the data well, i.e., the best-fit $\chi^2$ will be higher
than expected. In special cases, one or the other thing will happen. For instance, if
the effects on the CMB of the decaying neutrinos and their decay radiation is exactly
mimicked by some perturbation to the set of \cps, then a \lcdm model with offset
parameters will fit the data well. If, on the other hand, the effects of the decaying
neutrinos and the decay radiation are orthogonal to the effects of parameter offsets,
then the offsets will be small, but no model will fit the data well. If no \lcdm
model can reproduce a decaying neutrino model, in the sense that the best-fit
$\chi^2$ is large, then the decaying neutrino model is said to be distinguishable
from \lcdm.

If the offsets are small, then the problem can be analyzed analytically. This is done
in Appendix~\ref{ap:dist}. The procedure we use to determine the distinguishability
of a model is to calculate the $C_{l}$ spectrum and the Fisher matrix for the cosmic
parameters being considered, for the baseline \lcdm model. Then, for a given decaying
neutrino model we
\begin{itemize}
\item Find the parameter offsets using Eq.~\ref{eq:bias}.
\item Determine the probability distribution for the goodness of fit $\chi^2$. Being
approximately Gaussian, this distribution is fully characterized by expected the
best-fit $\left<\chi^2_{min}\right>$, given by Eq.~\ref{eq:chi2} and the variance
$\sigma_\chi$, given by Eq.~\ref{eq:sigma}.
\item Determine the level of distinguishability by convolving the
probability distribution for $\chi^2_{min}$ with the allowed level for each
$\chi^2_{min}$, as per Eq.~\ref{eq:Afinal}.
\end{itemize}

\section{Regions of Parameter Space}
\label{sc:regions}

In decaying neutrino scenarios, the physics of the neutrino decays, and therefore the
CMB anisotropy changes as the neutrino parameters are varied. It is therefore useful
to break the parameter space into regions and consider each region separately. To do
so we first note that several physical scales naturally divide the parameter space:
\begin{itemize}
\item $t_d = t_U$: This represents the division between stable and unstable
neutrinos. 
\item $\Omega_0 h^2 = 0.25$: For our decaying neutrino models, we let $\Omega_{CDM}$
vary to enforce a flat universe: $\Omega_0 = 1$. If the density in neutrinos or decay
radiation today is large enough, then $\Omega_0 > 1$ even with no CDM. For reasonable
values of $h_0$, regions with $\Omega_0 > 1$ tend to produce universes young enough
to violate independent age constraints: $\Omega_0 h^2 < 0.25$. For stable neutrinos,
this translates into the well-known bound on the sum of the masses, $\sum_i m_i \le
24$ eV, where the index $i$ runs over all neutrino species.
\item $t_d = t_{rec}$: The decay radiation for neutrinos that decay before
last scattering sources CMB anisotropy through early-ISW effect, while the decay
radiation for neutrinos that decay after last scattering creates a late-ISW effect.
\item $m_h = 3 T(t_d)$: This scale divides equilibrium ($m_h \lesssim 3T(t_d)$) from
out-of-equilibrium ($m_h \gtrsim 3T(t_d)$) decaying neutrinos. Neutrinos that decay
in equilibrium produce small changes in the radiation density, while those that decay
out-of-equilibrium produce larger effects.
\item $m_h = 3 T(t_{rec})$: This scale determines whether the decaying
neutrinos are relativistic ($m_h \lesssim 3 T(t_{rec}$) or non-relativistic ($m_h
\gtrsim 3 T(t_{rec}$) at last scattering.
\end{itemize}

Based on these scales, we have broken the decaying neutrino parameter space into
regions, labeled alphabetically, as shown in Fig.~\ref{fg:ps_labeled}:
\begin{figure}[!ht]
\centerline{\epsfig{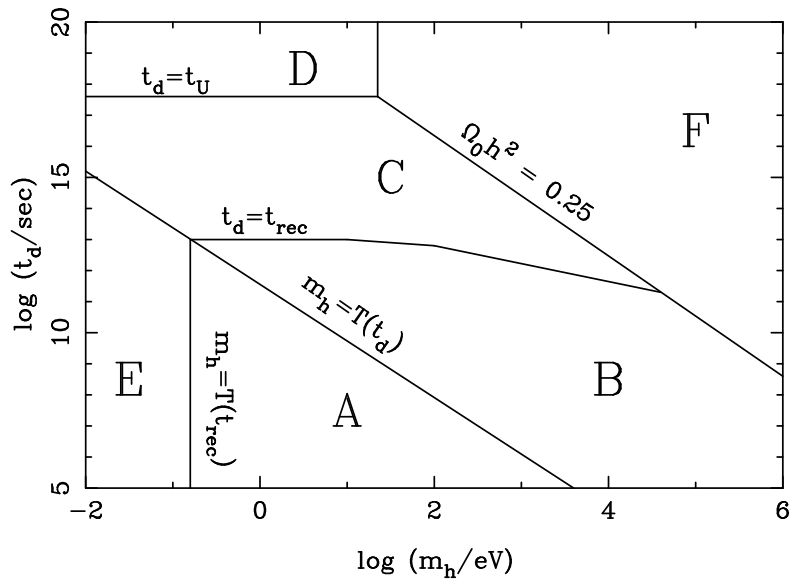}}
\caption{Decaying neutrino parameter space, divided into regions according to the
physics of the CMB anisotropy.}
\label{fg:ps_labeled}
\end{figure}

\begin{itemize}

\item {\em A}: $3 T(t_{rec}) < m_h < 3 T(t_d)$

In this region the neutrino decays in equilibrium, before last scattering. The energy
density in radiation is increased relative to the standard case by $\delta N_\nu =
0.17$. The only difference between the CMB anisotropy of these models and the
baseline model is due to this extra radiation. If the universe is not completely
matter-dominated at last scattering, then the gravitational potentials are decaying
at last scattering, when the primary anisotropy is being formed. Decaying potentials
at last scattering generate anisotropy through the early-ISW effect. The small amount
of extra radiation in these models induces a small amount of extra anisotropy. The
angular scale of the effect is determined by the sound horizon at last scattering,
placing the feature near the first acoustic peak, which, for the flat universes that
we consider, $l \sim 200$. The degeneracy in $m_h$ and $t_d$ means that these models
can be considered as a group.

Because the CMB anisotropy in this region depends only on the radiation density at
last scattering, the details of the decay channel are unimportant, except to the
extent that they determine this density. For example, it is easy to extend the
analysis to include the three-body decay scenario $\nu_h \rightarrow
\nu_l\,\nu_l\,\nu_l$, because we know that in this scenario, $\delta N_\nu = 0.52$.

The claim that we can calculate the CMB spectrum for models in region-$A$ by simply
adding $0.17$ species of massless neutrinos bears examination. One possible concern
follows from the fact that if massive neutrinos are present near last scattering,
then they will affect the CMB anisotropy. However, in this region there are no
massive neutrinos left at last scattering; they have decayed away by then. A more
serious concern involves spatial perturbations to the decay radiation. Treating the
decaying radiation by simply increasing the effective number of massless neutrino
species effectively assumes that the decay radiation perturbations are equal to
massless neutrino perturbations. But for times much later than those when decays are
important, the decay radiation perturbations approach those for massless
neutrinos. This is because the collision term in the Boltzmann equation that
describes the perturbation evolution, described in Sec.~\ref{sc:pertbe}, is only
important when decays are important, and the evolution equations without the
collision term are identical to those for standard massless neutrinos. In this
region, the decaying neutrinos decay away when they become non-relativistic, when
$T(t) \lesssim m_h/3$. If this time is much earlier than recombination, i.e., if $m_h
\ll 3 T(t_{rec})$, then the decay radiation perturbations can be approximated as
standard massless neutrinos, and the arguments in the last paragraph hold. From
Fig.~\ref{fg:ps_labeled}, this condition holds in region-$A$, for points a good deal
to the right of the defining line $m_h = 3 T(t_{rec})$. We will assume that this is
true for the rest of this work.

\item {\em B}: $m_h > 3 T(t_d)$, $t_d < t_{rec}$

Here, neutrinos decay out-of-equilibrium, before last scattering. Thus, as for
region-$A$, the decay radiation sources the early-ISW effect which results in extra
anisotropy near the first acoustic peak. But the effects are larger in this region
since out-of-equilibrium decays can generate large amounts of decay radiation, as
shown in Eq.~\ref{eq:nrd}. The amount of extra radiation, and hence the CMB
spectrum, depends on one parameter only, either $\alpha$ or $\delta N_\nu$. This is
in contrast this to the constant effect in region-$A$. Some models from region B,
parameterized by $\delta N_\nu$, are shown in Fig.~\ref{fg:cl_tcdm}. Another effect
is visible in addition to the early-ISW acoustic peak enhancement: a shift of all
features to smaller angular scales. This is due to the fact that, as the amount of
radiation at last scattering increases, the sound horizon at last scattering
decreases. For the standard $\Lambda$CDM model, the universe is mostly
matter-dominated at last scattering, with $\tau=2\sqrt{a}H_0^{-1}$. In the limit of a
completely radiation-dominated universe at last scattering, this relation is modified
to become $\tau=aH_0^{-1}$. This is the reason for the shift to smaller angular
scales, since $a$ at last scattering is the same in both scenarios.

\begin{figure}[!ht]
\centerline{\epsfig{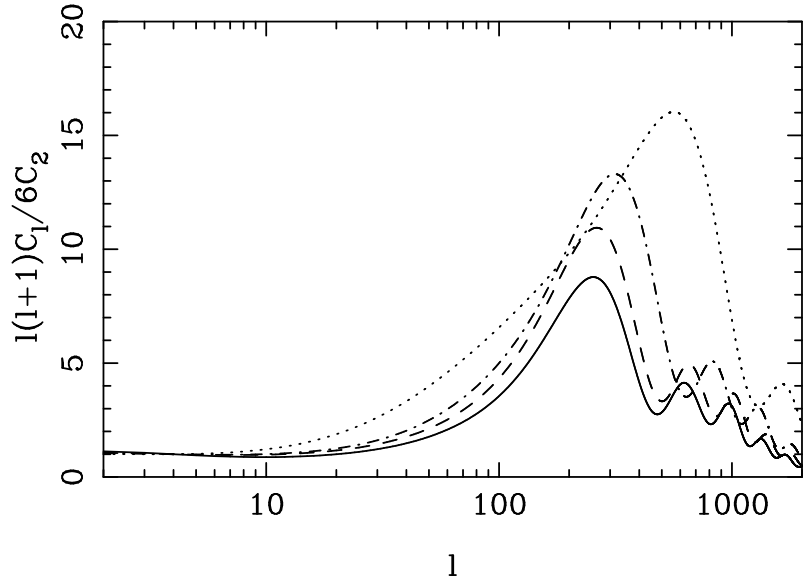}}
\caption{CMB anisotropies for early-decaying models, corresponding to region-$B$ of
parameter space. The quadrupole-normalized anisotropy is plotted as a function of
$l$. The solid line depicts \lcdm. The dashed line represents a model with $\delta
N_{rd} = 1.0$, the dashed-dotted has $\delta N_{rd} = 10.0$ and the dotted has
$\delta N_{rd} = 100.0$. Each value of $\delta N_{rd}$ corresponds to some $\alpha$
through Eq.~\ref{eq:nrd}, and thus to a one-parameter family of decaying neutrino
models. Note that as $\delta N_{rd}$ increase, the universe becomes less and less
matter-dominated at recombination and the early-ISW peak becomes more
prominent. For $\delta N_{rd} = 100$, the universe is very radiation-dominated at
last scattering. This changes the age of the universe at recombination and shifts
features to smaller values of $l$.}
\label{fg:cl_tcdm}
\end{figure}

In the future, we will parameterize models in region-$B$ in terms of the decay
radiation density $\delta N_\nu$. We should question the validity of this
parameterization. We would expect that the complicating effect from massive neutrinos
being present at last scattering is absent for $t_d \ll t_{ast}$, because in
region-$B$ the decaying neutrinos decay away when $t \sim t_d$. Furthermore, the
collision terms in the Boltzmann equations for the decay radiation vanish for $t \gg
t_d$, so that we expect that the decay radiation perturbations are well approximated
by massless neutrinos. In this region we have the advantage that we can check this
because we can calculate the CMB anisotropy properly. This is because region-$B$
the neutrinos decay out-of-equilibrium, where our Boltzmann hierarchy for the decay
radiation, Eq.~\ref{eq:drbe}, is valid. Because of this, it is possible to check the
accuracy of this approximation. Specifically, we have checked that the calculated CMB
spectrum in this region, for $t_d \ll t_{rec}$, is identical in the following two
approaches: 1) adding a separate Boltzmann hierarchy, described by Eq.~\ref{eq:drbe}
for the decay radiation perturbations, and 2) simply increasing the effective number
of massless neutrinos within a $\Lambda$CDM framework, using Eq.~\ref{eq:nrd}.

\item {\em C}: $m_h > 3 T(t_d)$, $t_{rec} < t_d < t_U$

In these models, the neutrinos decay out-of-equilibrium and after last
scattering. The decay radiation is not present until after last scattering; the
decays source anisotropy through the late-ISW effect. As for region-$B$, the amount
of decay radiation at decay is determined by the parameter $\alpha$. But for
region-$C$, the parameter degeneracy is broken, because the scale of the late-ISW
feature depends on the neutrino lifetime. The CMB spectra for several late-decaying
models are shown in Figs~\ref{fg:cl_t14},~\ref{fg:cl_t15},~\ref{fg:cl_t16}. Note that
the size of the ISW effect increases as $m_h$ increases, for fixed $t_d$, and the
location of the feature shifts to larger scales (smaller $l$) as $t_d$ increases.

\begin{figure}[!ht]
\centerline{\epsfig{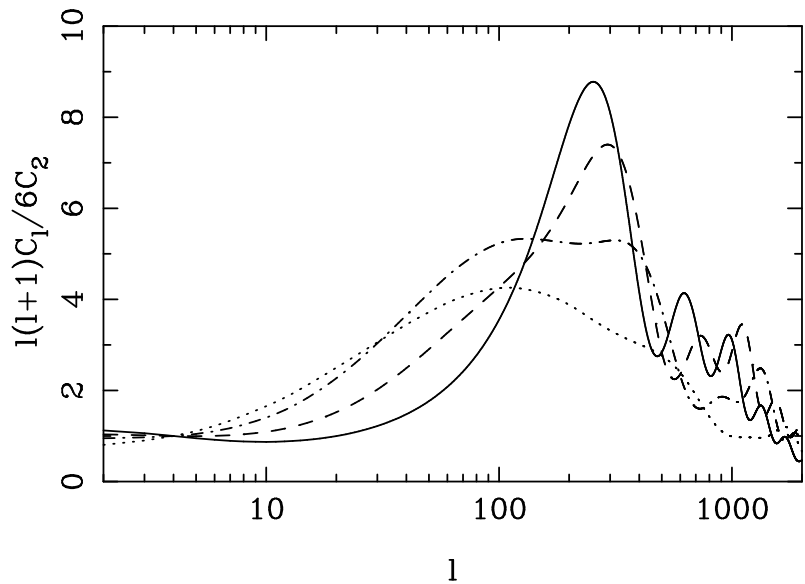}}
\caption{CMB anisotropies for late-decaying models with $t_d = 10^{14}$ sec. The
solid line represents the baseline \lcdm model. The dashed line has $m_h = 10$ eV,
the dashed-dotted line has $m_h = 31.4$ eV and the dotted has $m_h = 100$ eV. The
decay radiation sources a late-ISW feature that becomes more prominent for larger
masses.}
\label{fg:cl_t14}
\end{figure}

\begin{figure}[!ht]
\centerline{\epsfig{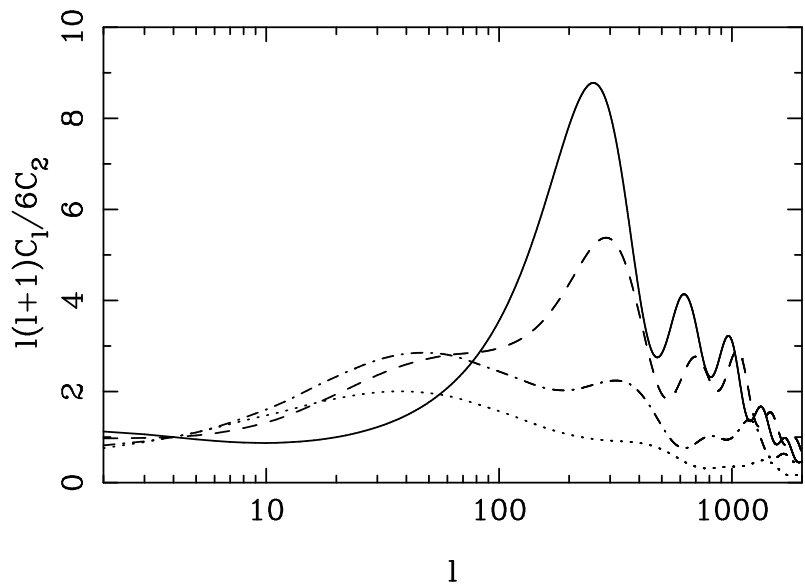}}
\caption{CMB anisotropies for late-decaying models with $t_d = 10^{15}$ sec. The
solid line represents the baseline \lcdm model. The dashed line has $m_h = 10$ eV,
the dashed-dotted line has $m_h = 31.4$ eV and the dotted has $m_h = 100$ eV. The
late-ISW feature is shifted to larger angles relative to the $t_d = 10^{14}$ sec
models.}
\label{fg:cl_t15}
\end{figure}

\begin{figure}[!ht]
\centerline{\epsfig{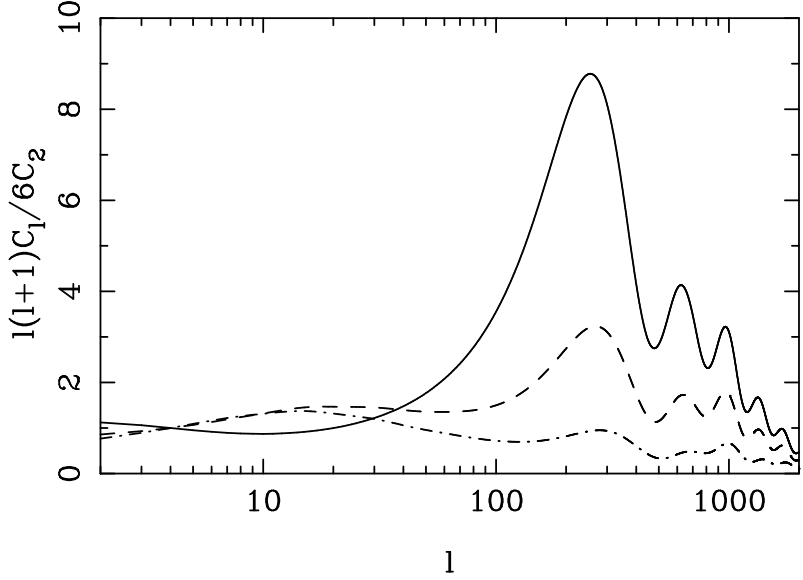}}
\caption{CMB anisotropies for late-decaying models with $t_d = 10^{15}$ sec. The
solid line represents the baseline \lcdm model. The dashed line has $m_h = 10$ eV,
the dashed-dotted line has $m_h = 31.4$ eV. The model with $m_h = 100$ eV is not
shown since this model has $\Omega_{rd} > 1 - \Omega_{B} - \Omega_\Lambda = 0.22$,
i.e., the model is in region-$E$. In these models, the late-ISW feature has
significant power at the quadrupole, which suppresses the small angle anisotropy in
this quadrupole-normalized plot. Of course, the normalization is allowed to vary in
all subsequent analysis.}
\label{fg:cl_t16}
\end{figure}

We can estimate the location of the late-ISW effect by noting that it is sensitive to
the scale of the sound horizon at the time the potentials decay. For neutrinos that
decay out-of-equilibrium, like those in regions-$B$ and $C$, this time is near $t =
t_d$, so that the location of the ISW induced feature is determined by the lifetime
of the neutrino. For lifetimes shorter than the age of the universe, inhomogeneities
on scales $k$ project onto angular scales $\ell \sim k\tau_0$ where $\tau_0$ is the
conformal time today (we assume a flat universe). The potentials vary in time, and
hence cause the ISW effect, most significantly at the time of decays on scales of
order the sound horizon: $k_{sh}^2 \simeq 3/(4\tau_d^2 w)$ where $w = P/\rho$ is the
averaged equation of state. Therefore, the bump in the spectrum is produced at $\ell
\sim k_{sh} \tau_0 \simeq (\tau_0/\tau_d)(4w/3)^{-1/2}$. If the decay occurs after
matter domination but before possible cosmological constant domination (which occurs
only at very late times), then $w$ is determined by the decay radiation. Since the
epoch of matter-radiation equality is near recombination for the models we are
considering, this assumption is valid for most of region-$C$. Hence $w
\simeq\Omega_{rd}(t_d)/3$, where $\Omega_{rd}(t_d)$ is the fraction of critical
density in decay radiation at decay. If we assume that the decay radiation never
dominates the universe, then we can estimate $\Omega_{rd}(t_d)$ in terms of the
neutrino properties:
\begin{equation}
\Omega_{rd}(t_d) \simeq 1.7\times 10^{-3} \, \frac{1}{h^2}
\left(\frac{m_h}{\rm{eV}}\right)^{4/3}
\,,
\end{equation}
valid for $\Omega_{rd} \ll 1$. Since we are assuming that the universe is matter
dominated at decay, physical times are related to conformal times by $\tau \propto
t^{1/3}$. If, on the other hand, $\Omega_{rd}(t_d) \simeq 1$, then the decay
radiation dominates until very late times, and we have the radiation-dominated
expression $\tau \propto t^{1/2}$. We can combine these results to obtain an
approximate expression for the location of the ISW peak for region-$C$:
\begin{equation}
l_{ISW} \simeq 
\left\{
\begin{array}{ll}
1200\,h\,\alpha^{-1/3} 
& \mbox{if $m_h \ll 120\,h^{3/2}$ eV}\\
1.5 \sqrt\frac{t_U}{t_d}
& \mbox{if $m_h \gtrsim 120\,h^{3/2}$ eV}
\end{array}
\right. \,,
\end{equation}
where $t_U \simeq 4\times 10^{17}$ sec is the age of the universe. Entropy
fluctuations, which occur when there are appreciable amounts of both matter and
radiation, decrease the sound speed, thereby increasing $l_{ISW}$. The relative size
of this effect is typically of order 20--40\pct.

\item {\em D}: $t_d > t_U$

In this region, the massive neutrino is effectively stable. Stable neutrinos have a
long history as a dark matter candidate. Constraints on these models have been
explored in Refs.~\cite{Dodelson96,Bond97}.

\item {\em E}: $m_h < 3 T(t_d)$, $m_h < 3 T(t_{rec})$

Here the neutrinos decay in equilibrium. Therefore, the energy density in radiation
increases by $\delta N_\nu = 0.17$ after the neutrino becomes non-relativistic and
decays away. However, since $m_h \lesssim 3 T(\tau_\ast)$, this occurs after last
scattering, with the exact time depending on $m_h$; the CMB anisotropy in this
region are degenerate in $\tau_d$. The small late-ISW effect that is induced is too
small to be measured, even with future satellite-based experiments. For this reason,
we will not study region-$E$ any further.

\item {\em F}: $m_h < T(t_d)$, $t_d > t_{rec}$

Here, the density in either stable neutrinos or their decay radiation is enough to
require $\Omega_0 > 1$. These models are extreme and suffer several problems, such as
producing a universe that is too young, and so will not be analyzed further here.

\end{itemize}

\section{Results}
\label{sc:results}

The goal of this section is to answer two questions. 1) Is the CMB anisotropy for
some decaying neutrino model sufficiently different from baseline \lcdm so that the
two models are distinguishable? 2) Given a particular decaying neutrino model, how
well can the cosmic parameters, including neutrino parameters, be measured? To answer
question 1) we use the distinguishability framework of Sec.~\ref{sc:dist} and the
Appendix, and to answer question 2) we use the Fisher matrix approach of
Sec.~\ref{sc:fim}.

\begin{figure}[!ht]
\centerline{\epsfig{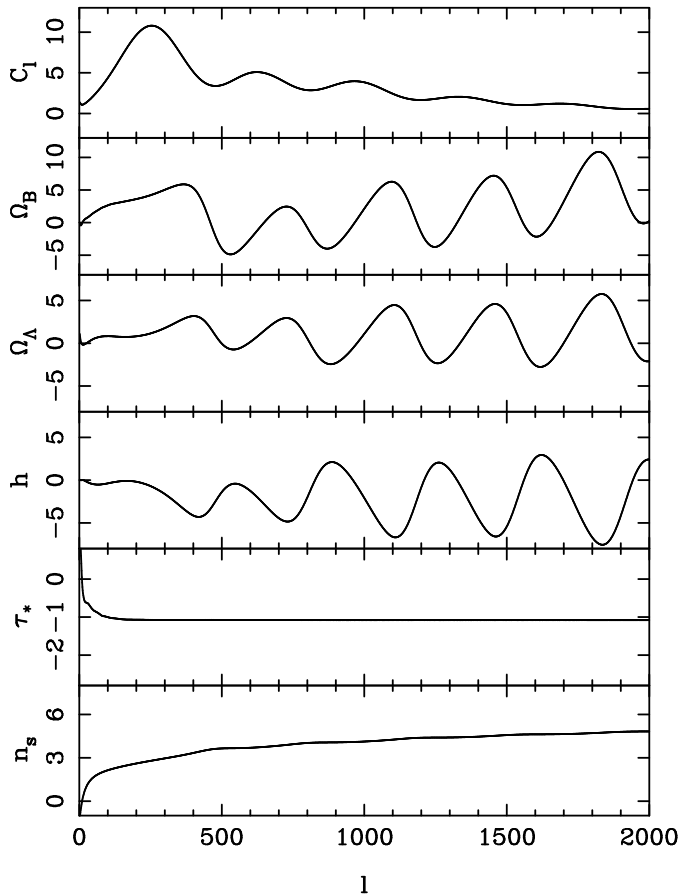}}
\caption{Baseline \lcdm model and its derivatives with respect to cosmic
parameters. The top panel is the quadrupole-normalized baseline CMB spectrum:
$\Omega_B = 0.08$, $\Omega_\Lambda = 0.7$, $h = 0.5$, $\tau_\ast = 0.1$, $n_s =
1.0$. The lower panels are derivatives with respect to $\Omega_B$, $\Omega_\Lambda$,
$h$, $n_s$, and $\tau_\ast$, normalized to the baseline spectrum: $1/C_{Xl} \; \partial
C_{Xl} / \partial \lambda_i$. The derivative with respect to $Q$ is not shown since
$\partial C_{Xl}/\partial Q \propto C_{Xl}$.}
\label{fg:base_deriv}
\end{figure}

In both cases, we adopt the following \lcdm model as our baseline: $\Omega_\Lambda =
0.7$, $\Omega_{CDM} = 0.22$, $\Omega_B = 0.08$, $h = 0.5$, Harrison-Zeldovich
primordial spectrum ($n_s = 1.0$), reionization optical depth $\tau_\ast = 0.1$, and
three massless species of neutrinos. The set of cosmic parameters allowed to vary was
$\lambda_i = \{\Omega_\Lambda,\,\Omega_B,\,h,\,n_s,\,\tau_\ast,\,Q\}$, where $Q$ is
the overall normalization. To calculate the Fisher matrix $\alpha_{ij}$ we took
two-sided derivatives for all of our cosmic parameters parameters as suggested in
reference~\cite{Eisenstein98}, i.e.,
\begin{equation}
\frac{\partial C_{Xl}}{\partial \lambda_i} = 
  \frac{C_{Xl}(\lambda_i + \delta \lambda_i) - 
    C_{Xl}(\lambda_i - \delta \lambda_i)}
    {2\delta \lambda_i}
\,,
\end{equation}
where $\delta \lambda_i$ is the numerical stepsize in the $i$-th cosmic
parameter. All of our derivative stepsizes were taken to be 3\pct of their baseline
values, except for $\tau_\ast$, whose stepsize was $0.03$. We verified numerically
that the derivatives were stable with respect to varying the stepsizes. In
calculating $\partial C_{Xl} / \partial \Omega_B$ and $\partial C_{Xl} /
\partial \Omega_\Lambda$, we allowed $\Omega_{CDM}$ to vary, to maintain flat
universe: $\Omega_{CDM} = 1 - \Omega_\Lambda - \Omega_B$. The baseline model and its
derivatives are shown in Fig.~\ref{fg:base_deriv}.

From the CMB spectrum and its derivatives, we calculated the Fisher matrix, using
Eq.~\ref{eq:fm}. To analyze a real experiment requires understanding details like
their window functions and experimental noise. However, for future satellite-based
experiments like MAP and Planck, the experimental uncertainty is expected to be below
cosmic variance for most of the angular scales they are designed to measure, and the
window functions are relatively narrow. This allows us to characterize the
experiments as cosmic variance limited to some $l_{max}$, with the value of $l_{max}$
determined by the experiment. We take $l_{max} = 1000$ for MAP and $l_{max} = 2500$
for Planck. For both values of $l_{max}$ we consider cases with and without
polarization information. The reason for this is that it is not certain how good
polarization information will be. For the case that includes polarization, we assume
cosmic-variance limited polarization information from a minimum $l_{min} = 200$ up to
the same $l_{max}$ as for the temperature data. The reason for the minimum value of
$l$ is that the large-scale polarization signal is small enough to be overwhelmed by
the experimental noise of MAP and Planck. Our results are insensitive to the precise
value of $l_{min}$. The statistical uncertainties on the cosmic parameters are shown
for MAP and Planck in Tab.~\ref{tb:stat_base}. In this work, we take the conservative
(and realistic) approach of always marginalizing over all cosmic parameters
simultaneously. In this case, Eq.~\ref{eq:delcp} gives the statistical uncertainty on
the parameters.

\begin{table}
\begin{center}
\begin{tabular}{|c|c|c|c|c|} \hline
Parameter & $l_{max} = 1000$ & $l_{max} = 2500$  &
  $l_{max} = 1000$ (w/ pol.) & 
  $l_{max} = 2500$ (w/ pol.) \\
\hline 
$\Omega_B$ & 5.16 \pct & 2.01 \pct & 2.27 \pct & 0.700 \pct \\ 
$\Omega_\Lambda$ & 4.02 \pct & 1.26  \pct & 1.71 \pct & 0.391 \pct \\ 
$h$ & 3.56 \pct & 1.05 \pct & 1.46 \pct & 0.323 \pct \\ 
$n_s$ & 1.46 \pct & 0.340 \pct & 0.665 \pct & 0.194 \pct \\ 
$\tau$ & 0.0803 & 0.0523 & 0.0579 & 0.0507 \\
$Q$ & 5.75 \pct & 5.29 \pct & 5.37 \pct & 5.24 \pct \\ 
\hline
\end{tabular}
\end{center}
\label{tb:stat_base}
\caption{Statistical uncertainties on cosmic parameters for the fiducial \lcdm model
for $l_{max} = 1000$ and for $l_{max} = 2500$, with and without polarization
information. In all cases all cosmic parameters were allowed to vary simultaneously.}
\end{table}

\subsection{Ruling out models}

\begin{figure}[!ht]
\centerline{\epsfig{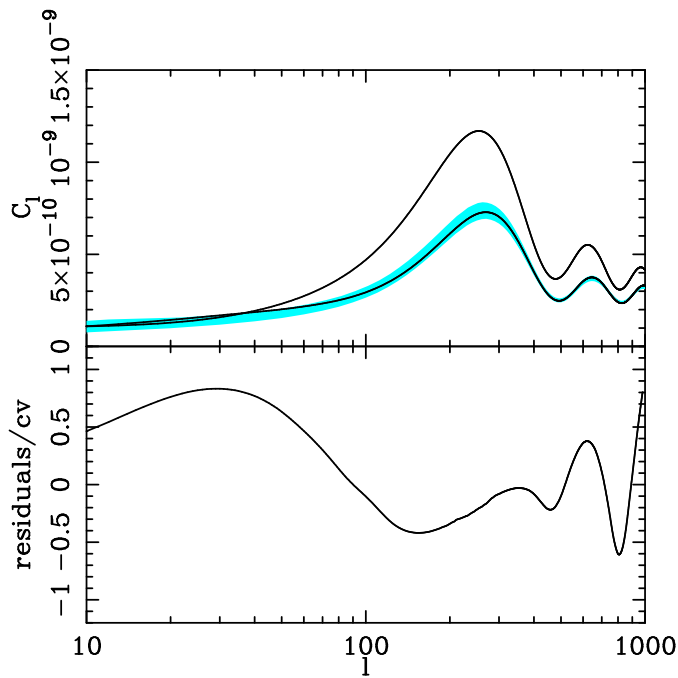}}
\caption{Example of distinguishability analysis for a late-decaying neutrino
scenario. Here $m_h = 3.23$ eV, and $t_d = 10^{16}$ sec. In the top panel the dashed
line depicts the baseline $\Lambda$CDM model, with arbitrary normalization. The solid
black line shows the decaying neutrino spectrum and the grey line shows the best-fit
perturbed $\Lambda$CDM model. The thickness of the grey line represents cosmic
variance. The bottom panel shows the difference between the decaying neutrino
spectrum and the best-fit perturbed $\Lambda$CDM model, in units of cosmic
variance. This model produces an ISW peak near $l = 25$, whose signature can clearly
be seen in the bottom pane - no values of cosmic parameters in a $\Lambda$CDM model
can reproduce such a feature. For MAP, without polarization, this model is ruled out
at the 89.7 \pct level.}
\label{fg:lateDecayDist}
\end{figure}

\begin{figure}[!ht]
\centerline{\epsfig{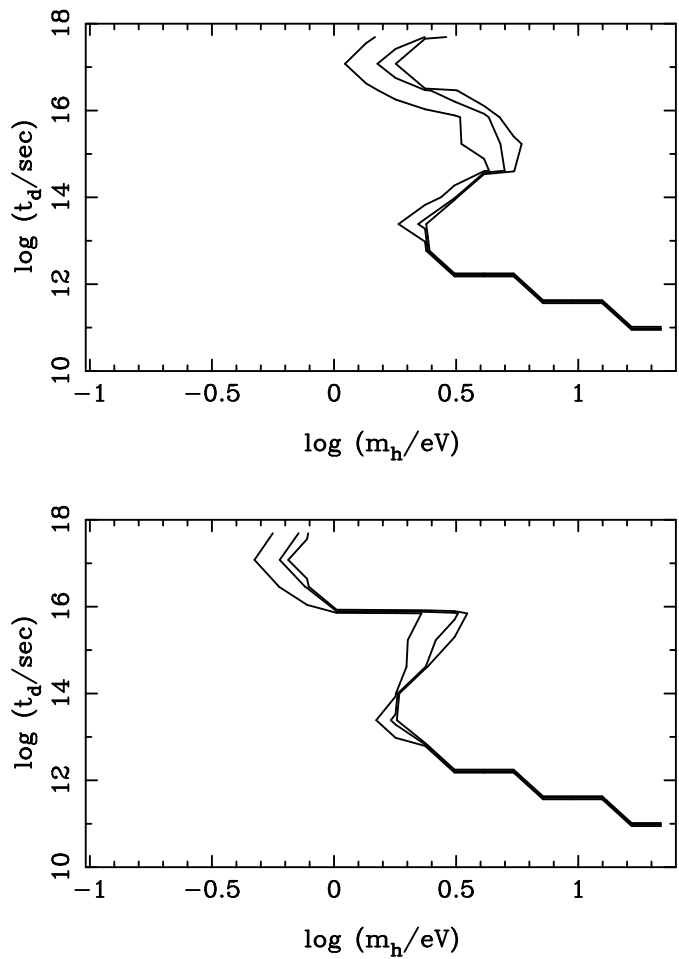}}
\caption{Decaying neutrino parameter space, showing models that are distinguishable
from \lcdm. The three contours represent distinguishability at the 90\pct, 99.9\pct
and 99.9\pct levels for the MAP experiment. In the top panel, temperature data is
considered alone; the bottom panel includes polarization. Models to the right of the
contours are distinguishable.}
\label{fg:contourMap}
\end{figure}

\begin{figure}[!ht]
\centerline{\epsfig{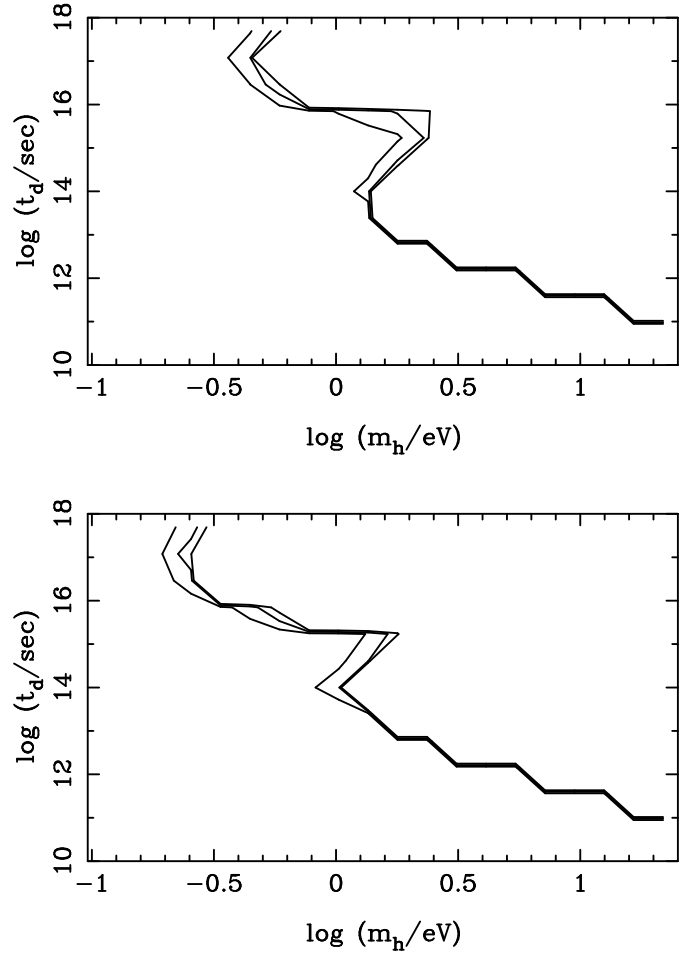}}
\caption{Same as Fig.~\ref{fg:contourMap}, but for the Planck experiment.}
\label{fg:contourPlanck}
\end{figure}

We analyzed a grid of models, consisting of 20 masses with $\log(m_h/\rm{eV})$ evenly
spaced from -1.0 to 1.40, and 13 lifetimes with $\log(t_d/\rm{sec})$ evenly spaced
from 10.0 to 18.0. For each grid point, we followed the procedure given in
Sec.~\ref{sc:dist} and the Appendix. An example of this procedure, for a
late-decaying scenario, is shown in Fig.~\ref{fg:lateDecayDist}. There we show the
\lcdm and decaying neutrino spectrum, along with the best-fit perturbed
\lcdm model and the discrepancy in the fit in units of cosmic variance. From
this discrepancy we calculate a confidence level for the model. The results for MAP
and Planck are shown in Figs.~\ref{fg:contourMap} and \ref{fg:contourPlanck}. For
MAP, stable neutrinos of masses greater than a couple of eV are distinguishable from
the baseline model, while for Planck, the sensitivity extends down to masses of
several tenths of an eV. As the lifetime decreases and the neutrino becomes unstable,
but late-decaying, the sensitivity in mass decreases somewhat. This is because the
late-ISW signature of a late-decaying neutrino is mostly degenerate with
reionization. When the lifetime is short enough so that the neutrinos are decaying
before last scattering, models with the same value of $\alpha$ are degenerate, and
are distinguished at the same level. This is clear from a visual inspection of the
plot. Finally, even the most optimistic case of Planck with polarization will not be
able to distinguish equilibrium decaying models from
\lcdm.

\begin{figure}[!ht]
\centerline{\epsfig{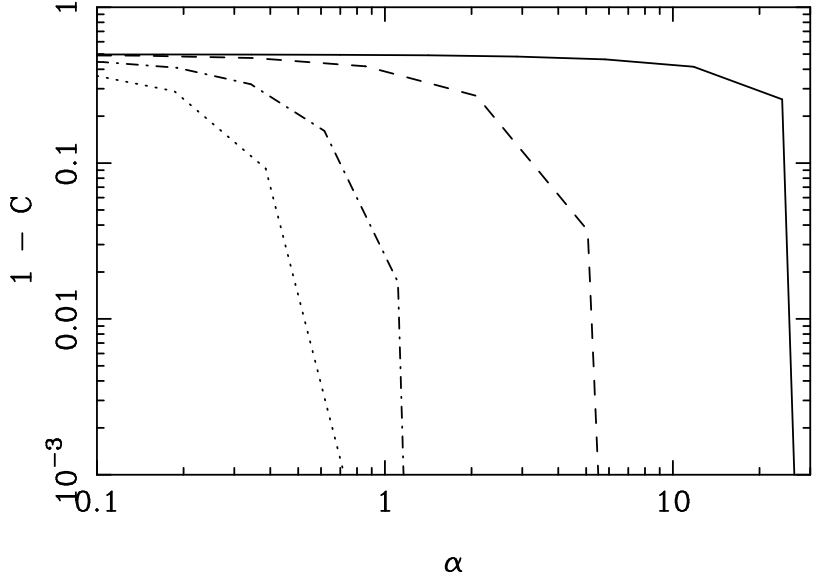}}
\caption{Level at which early-decaying models are allowed, as a function of
relativity parameter $\alpha = 0.087 (m_h/\rm{MeV})^2 (t_d/\rm{sec})$, for $l_{max} =
$ $1000$ without polarization (solid line), $1000$ with polarization (dashed line),
$2500$ without polarization (dash-dotted line) and $2500$ with polarization (dotted
line). Equilibrium-decaying neutrinos, corresponding to region-$A$, have $\alpha <
1$, whereas neutrinos that decay out-of-equilibrium, region-$B$, have $\alpha >
1$. For two-body decays, $\delta N_\nu
\rightarrow 0.17$ as $\alpha \rightarrow 0$. An experiment sensitive to $l_{max} =
2500$, with or without polarization information, will be sensitive to neutrinos on
the border between equilibrium and out-of-equilibrium decay ($\alpha \sim
1$). Without polarization, an experiment sensitive only to $l_{max} = 1000$ can only
probe very out-of-equilibrium decays ($\alpha > 100$); including polarization
increases the sensitivity to $\alpha \simeq 5$. None of the cases considered will be
able to distinguish equilibrium-decaying models from \lcdm.}
\label{fg:aVsAlpha}
\end{figure}

For early-decaying neutrinos, we can obtain a clearer picture by exploiting the
parameter degeneracy, describing the models with the single variable
$\alpha$. Fig.~\ref{fg:aVsAlpha} shows the confidence level for models as a function
of $\alpha$. MAP will be able to distinguish models with $\alpha \gtrsim 10$ without
polarization, and $\alpha \gtrsim 5$ with polarization. Planck, with or without
polarization, will distinguish any out-of-equilibrium decaying models, with $\alpha
\gtrsim 1$. This plot confirms the result that models in region-$A$, with $\delta
N_\nu = 0.17$ at recombination, are indistinguishable from \lcdm. 

The formalism used to perform these distinguishability calculations is valid in a
linear regime, where Eq.~\ref{eq:smalldc} holds. If the parameter biases become
large then the formalism breaks down. Since some of the decaying neutrino models
produce CMB anisotropy very different from the canonical $\Lambda$CDM, the linear
approximation must break down for these models. However, the distinguishability
contours can be believed if two facts hold. First, the linear approximation should
hold for models that are just becoming indistinguishable, i.e., those along the
contour lines in Figs.~\ref{fg:contourMap} and \ref{fg:contourPlanck}. Second, models
that are inside the contour must stay indistinguishable. The first point we observe
to be true numerically. The second point could break down in a couple of ways: a) the
spectra start to look more like standard $\Lambda$CDM as we go inside a contour, or
b) the spectra don't look like our baseline $\Lambda$CDM but instead look like some
standard model with very perturbed parameters. Neither objection holds. The first is
obviously false because for any fixed $t_d$, the decaying neutrino effects increase
as we go inside the contour, increasing $m_h$. The second objection is only slightly
more problematic. For late decaying neutrinos, the decaying neutrino feature is a
late-ISW bump at some large angular scale - it's pretty easy to see that this cannot
be mimicked by $\Lambda$CDM with perturbed cosmic parameters. For early decays, the
early-ISW effect is degenerate with the ratio of matter density to radiation density
at last scattering. But in this work the only non-standard physics we are allowing is
the decaying neutrino itself. This allows us to fix this ratio for the set of
$\Lambda$CDM models. Other cosmic parameters affect the relative amount of radiation
at last scattering. In particular $h$, and $\Omega_\Lambda$, are mostly degenerate
with $N_\nu$~\cite{Lopez98}. However, the degeneracy is not complete so that $h$ and
$\Omega_\Lambda$ cannot completely mimic the early-ISW signal for these models. Since
we are considering models well within the distinguishability contours where the
early-ISW signal is large, the lack of complete degeneracy prevents $h$ and
$\Omega_\Lambda$ from mimicking the decaying neutrino signal.

\subsection{Measuring neutrino parameters}

Here we are concerned with our ability to measure cosmic parameters, where the set
includes quantities that specify the decaying neutrinos. We are primarily interested
in the answers to two questions. First, what are the statistical uncertainties in the
neutrino parameters? This goes to the goal of using the CMB as a probe of neutrino
physics. Second, how much are the uncertainties in the non-neutrino cosmic parameters
degraded by their presence? It is always true that adding extra parameters to the set
increases or at best doesn't change the uncertainty in the existing parameters. If
the extra parameters are orthogonal to the existing parameters in the sense that the
change in the CMB spectrum from perturbing the new parameters cannot be mimicked by
perturbing the existing parameters, then the degradation in the existing
uncertainties is minimal. If, in the other extreme, the effect of perturbing new
parameters can be mimicked by changing the existing parameters, the degradation is
severe. Mathematically, this can be analyzed in terms of cross-correlations in the
Fisher matrix: large cross-correlations mean degraded sensitivities. This degradation
is one of the main arguments for pursuing the distinguishability calculations of the
last section. If the CMB provides no evidence for decaying neutrinos, i.e., the
real-universe CMB spectrum is not distinguishable from \lcdm, then adding decaying
neutrino parameters will be a hard sell.

Since the physics behind the CMB anisotropy is different for different regions of
neutrino parameter space, there is no one best set of neutrino parameters to add to
the cosmic parameters. We will group together the early-decaying neutrino models,
corresponding to region-$A$ and region-$B$, together, and use $\alpha$ as the sole
neutrino parameter in this region. It is easier to compute the CMB spectra in terms
of the radiation energy density $N_\nu$, but $\alpha$ is more directly related to the
mass and lifetime of the neutrino; the uncertainty in $\alpha$ can be related to the
uncertainty in $N_\nu$ through the relation
\begin{equation}
\delta \alpha_{stat} = \delta N_{\nu,stat} 
  \frac{\partial \alpha}{\partial N_{\nu}}
\,,
\end{equation}
where the derivative is obtained from a numerical solution to the Boltzmann equation,
summarized in Fig.~\ref{fg:dn_a}. For the late-decaying, but unstable neutrinos in
region-$C$, we can just use the mass and lifetime as our neutrino parameters,
$(m_h,t_d)$. However, as the neutrinos become stable, in region-$D$, the inverse of
the lifetime becomes a more natural parameter, since the baseline model corresponds
to the limit $m_h \rightarrow 0$ and $1/t_d \rightarrow 0$. Therefore, we define an
inverse-lifetime parameter, scaled to the lifetime of the universe,
\begin{equation}
y \equiv \frac{t_U}{t_d}
\,,
\end{equation}
and use the set $(m_h,y)$ in region-$D$. Our parameter choices are summarized in
Tab.~\ref{tb:regions}.

\begin{table}
\begin{center}
\begin{tabular}{|c|c|} \hline
Region & Neutrino Parameter(s)
\\ \hline
$A$ & $\alpha$ (degenerate, with $\delta N_\nu = 0.17$) \\
$B$ & $\alpha$ \\
$C$ & $(m_h,\,t_d)$ \\
$D$ & $(m_h,\, y)$ \\
\hline
\end{tabular}
\end{center}
\caption{Neutrino parameters to add to the set of cosmic parameters, for different
regions of neutrino parameter space.}
\label{tb:regions}
\end{table}

\begin{figure}[!ht]
\centerline{\epsfig{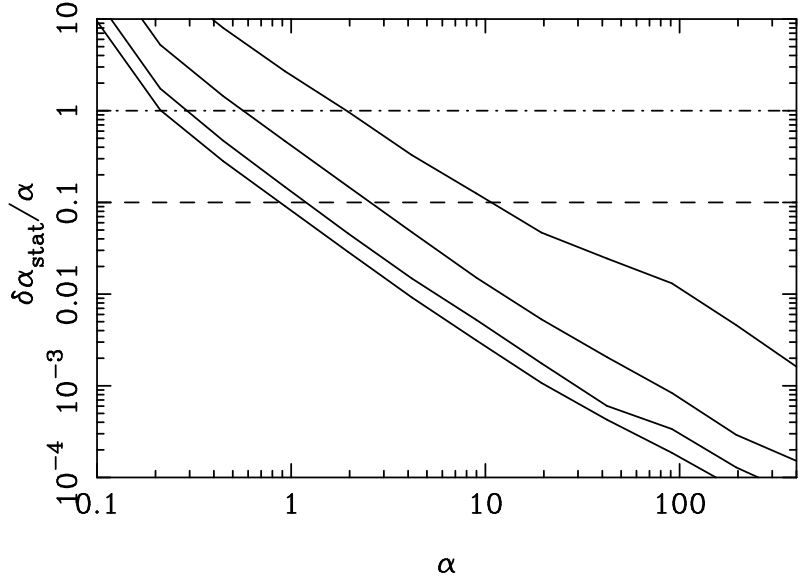}}
\caption{Using the CMB to measure $\alpha$ for early-decaying neutrinos,
corresponding to regions-$A$ and $B$. The solid lines show the statistical
uncertainty in the parameter $\alpha$ as a function of $\alpha$. In order of
increasing sensitivity, the solid lines correspond to $l_{max} = $ 1000 (no
polarization), $l_{max} = 1000$ (with polarization), $l_{max} = 2500$ (no
polarization) and $l_{max} = 2500$ (with polarization). The dot-dashed line
represents the case where the $\delta \alpha_{stat} = \alpha$; the dashed line shows
$\delta \alpha_{stat} = 0.1 \alpha$. For models below these lines, $\alpha$ can be
measured to good relative accuracy.}
\label{fg:del_alpha_tcdm}
\end{figure}

Fig.~\ref{fg:del_alpha_tcdm} shows the relative statistical uncertainty in $\alpha$,
versus $\alpha$, for early-decaying models. Note that $\delta \alpha_{stat} / \alpha$
is a decreasing function of $\alpha$. This is because the relative sensitivity to
radiation energy density is roughly model-independent, i.e., $\delta N_{\nu,stat} /
N_\nu$ is roughly constant. This means that $\delta \alpha_{stat} /
\alpha \propto \alpha^{-2/3}$, for $\alpha \gg 1$. For $\alpha \ll 1$, the radiation
density ceases to depend on $\alpha$ at all: $\delta \alpha_{stat}/\alpha \rightarrow
\infty$ as $\alpha \rightarrow 0$. The value of $\alpha$ with $\delta \alpha_{stat} =
\alpha$ is interesting because there \lcdm, with $\alpha = 0$ is ruled out at the
1-$\sigma$ level. For MAP, this point occurs at $\alpha \simeq 25$ without and
$\alpha \simeq 5$ with polarization. For Planck, with or without polarization, this
occurs near $\alpha = 1$. Models where the data would rule out \lcdm at high
significance occur for only slightly higher values of $\alpha$. These values should
be compared to the distinguishability results from the last section, where
distinguishable values of $\alpha$ were a factor of several higher. This represents
the advantage of including neutrino parameters in the analysis: one can rule out more
models this way.

\begin{figure}[!ht]
\centerline{\epsfig{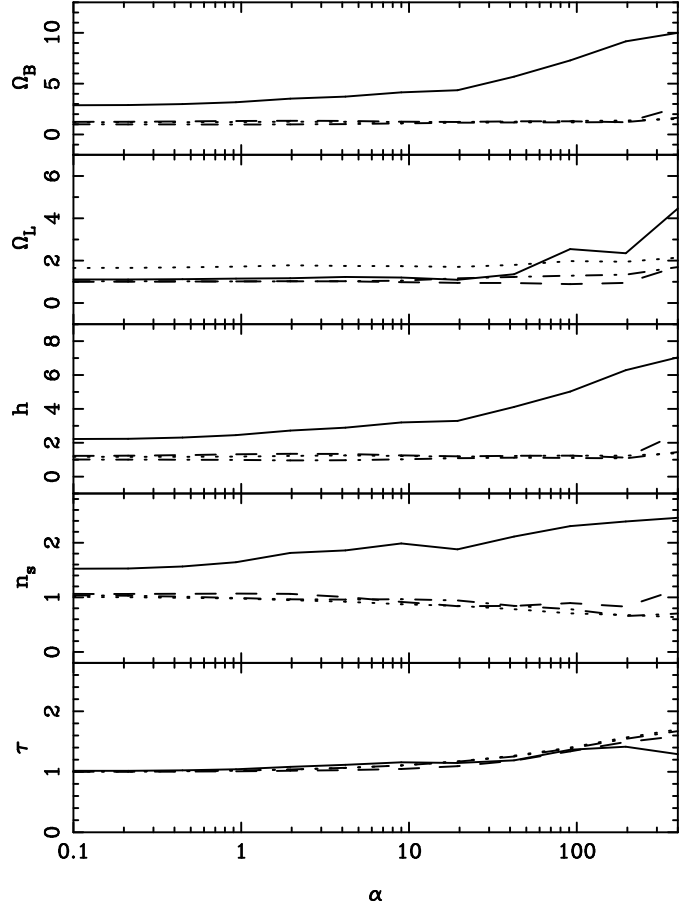}}
\caption{Degradation in ability to measure the non-neutrino parameters, for
early-decaying models. Each panel shows the statistical uncertainty in a cosmic
parameter as a function of $\alpha$. The uncertainties are normalized to the value
obtained analyzing \lcdm without decaying neutrino parameters. The different curves
in each panel correspond to MAP without polarization (solid), MAP with polarization
(long-dash), Planck without polarization (dash-dot), and Planck with polarization
(dotted). As $\alpha \rightarrow 0$, the CMB anisotropy is close enough to \lcdm so
that in this limit the curves may be interpreted as the degradation caused by adding
$\alpha$ as a parameter to \lcdm.}
\label{fg:delp_tcdm}
\end{figure}

This advantage comes at a price: the degradation in the ability to measure the
non-neutrino parameters. First, consider $\alpha \lesssim 1$, where the decaying
neutrino models produce CMB spectra very similar to \lcdm,and the uncertainties in
the \cps reflect the degradation that would occur if one added $\alpha$ (or $\delta
N_\nu$), and analyzed . The results for the limit $\alpha \rightarrow 0$ are shown in
Tab~\ref{tb:degr}. Note that the {\em relative} degradation is much larger for MAP
than for Planck, and that the degradation for certain parameters, $\Omega_B$, $h$,
$n_s$, is quite severe. Fig.~\ref{fg:delp_tcdm}, shows the statistical uncertainties
in the other cosmic parameters as a function of $\alpha$, also normalized to the
uncertainties for \lcdm. The table just discussed is the $\alpha \rightarrow 0$ limit
of the figure. The most prominent feature of the figure is general trend towards
lower sensitivity for increasing $\alpha$.

\begin{table}
\begin{center}
\begin{tabular}{|c|c|c|c|c|} \hline
Parameter & $l_{max} = 1000$ & $l_{max} = 2500$ &
$l_{max} = 1000$ (polarization) & $l_{max} = 2500$ (polarization) \\ 
\hline 
$\Omega_B$ 
& 2.85 & 1.00 & 1.23 & 1.21 \\ 
$\Omega_\Lambda$ 
& 1.11 & 1.03 & 1.00 & 1.65 \\ 
$h$ 
& 2.21 & 1.01 & 1.21 & 1.18 \\ 
$n_s$ 
& 1.52 & 1.04 & 1.06 & 1.02 \\ 
$\tau$ 
& 1.02 & 1.00 & 1.00 & 1.00 \\
$Q$ 
& 1.01 & 1.00 & 1.00 & 1.00 \\ 
\hline
\end{tabular}
\end{center}
\label{tb:degr}
\caption{Statistical uncertainties in non-neutrino parameters for $\alpha = 0$, i.e.,
where $\alpha$ is added as a cosmic parameter. The results are shown normalized to
the case where the data is analyzed without neutrino parameters. Therefore, the
numbers represent the degradation in sensitivity from including non-neutrino
parameters. Results are shown for $l_{max} = 1000$, and 2500, with and without
polarization.}
\end{table}

The parameter uncertainties for several late-decaying models in region-$C$ are shown
in Tables~\ref{tb:stat_late} and \ref{tb:stat_late_pol}. The uncertainty in the
neutrino parameters $m_h$ and $t_d$ increases as the lifetime increases from
$10^{14}$ to $10^{16}$ sec. The reason for this is that the ISW peak for the lower
lifetime neutrinos occurs at higher $l$, where two features work to improve the
sensitivity. First, cosmic variance is lower. Second, a given range of angular scales
translates to a larger number of $l$'s. As for the other parameters, since we add two
extra parameters to the analysis implies that we might expect relatively large
uncertainties. This is observed for most parameters, especially for MAP. For some
cases, the uncertainties are actually less than for \lcdm, which appears to violate
the requirement that adding \cps\xspace {\em decreases} the sensitivity in the other
parameters. However, here it doesn't make sense to talk about degradation, since the
CMB spectra for these models is very different from (these models are all
distinguishable - see Figs.~\ref{fg:contourMap} and \ref{fg:contourPlanck}).

\begin{table}
\begin{center}
\begin{tabular}{|c|c|c|c|c|c|c|c|c|} \hline
Model & $m_h$ & $t_d$ & $\Omega_B$ & $\Omega_\Lambda$ & $h$ & $n_s$ &
$\tau_\ast$ & $Q$ 
\\ \hline
$m_h = 10$ eV, $t_d = 10^{14}$ sec & 
2.15 & 8.05 & 26.1 & 16.1 & 14.1 & 2.67 & 83.4 & 6.77 \\
& 
& & (5.06) & (4.00) & (3.97) & (1.82) & (1.04) & (1.18) \\
&&&&&&&&\\
& 
1.21 & 5.59 & 4.12 & 2.34 & 2.05 & 0.403 & 82.2 & 5.40 \\
&
& & (1.81) & (1.36) & (1.40) & (0.60) & (1.42) & (1.00) \\ 
\hline
$m_h = 10$ eV, $t_d = 10^{15}$ sec & 
2.46 & 13.1 & 33.2 & 22.1 & 18.6 & 3.44 & 49.6 & 7.79 \\
& 
& & (6.44) & (5.51) & (5.22) & (2.35) & (0.62) & (1.35) \\
&&&&&&&&\\
& 
1.12 & 11.8 & 4.83 & 3.08 & 2.53 & 0.436 & 45.4 & 7.51 \\
&
& & (2.13) & (1.80) & (1.73) & (0.656) & (0.784) & (1.40) \\ 
\hline
$m_h = 10$ eV, $t_d = 10^{16}$ sec & 
15.6 & 54.4 & 28.2 & 21.5 & 15.2 & 2.57 & 122. & 7.20 \\
& 
& & (4.46) & (5.34) & (4.27) & (1.75) & (1.52) & (1.25) \\
&&&&&&&&\\
& 
11.4 & 44.9 & 9.46 & 8.96 & 4.68 & 0.827 & 90.7 & 6.69 \\
&
& & (4.17) & (5.23) & (3.20) & (1.24) & (1.57) & (1.25) \\ 
\hline
$m_h = 3.16$ eV, $t_d = 10^{14}$ sec &
6.18 & 25.7 & 13.8 & 8.53 & 7.63 & 1.77 & 105. & 6.98 \\
& 
& & (2.68) & (2.12) & (2.14) & (1.21) & (1.31) & (1.21) \\
&&&&&&&&\\
& 
3.14 & 15.3 & 5.04 & 2.82 & 2.53 & 0.654 & 90.2 & 6.53 \\
&
& & (2.22) & (1.65) & (1.73) & (0.983) & (1.56) & (1.21) \\ 
\hline
$m_h = 3.16$ eV, $t_d = 10^{15}$ sec & 
6.91 & 43.6 & 13.1 & 8.73 & 7.65 & 1.52 & 71.8 & 7.10 \\
& 
& & (2.53) & (2.17) & (2.15) & (1.04) & (.894) & (1.23) \\
&&&&&&&&\\
& 
4.02 & 36.7 & 4.90 & 2.98 & 2.56 & 0.537 & 70.1 & 6.62 \\
&
& & (2.16) & (1.74) & (1.76) & (0.807) & (1.21) & (1.23) \\ 
\hline
$m_h = 3.16$ eV, $t_d = 10^{16}$ sec &
23.0 & 119. & 18.3 & 10.9 & 10.1 & 1.74 & 146. & 7.94 \\
& 
& & (3.54) & (2.72) & (2.83) & (1.19) & (1.82) & (1.38) \\
&&&&&&&&\\
& 
12.6 & 50.3 & 6.28 & 4.35 & 3.28 & 0.675 & 80.5 & 7.12 \\
&
& & (2.77) & (2.53) & (2.24) & (1.02) & (1.39) & (1.33) \\ 
\hline
\end{tabular}
\end{center}
\label{tb:stat_late}
\caption{Using the MAP experiment to measure $m_h$ and $t_d$ for late-decaying
neutrinos. The statistical uncertainties on the cosmic parameters, $\delta
\lambda_i / \lambda_i$, in percent, are shown for several models. The number in
parenthesis is the ratio of the uncertainty to the uncertainty for \lcdm. For each
model the top row of data is for temperature data only and the bottom row includes
polarization.}
\end{table}

\begin{table}
\begin{center}
\begin{tabular}{|c|c|c|c|c|c|c|c|c|} \hline
Model & $m_h$ & $t_d$ & $\Omega_B$ & $\Omega_\Lambda$ & $h$ & $n_s$ &
$\tau_\ast$ & $Q$ 
\\ \hline
$m_h = 10$ eV, $t_d = 10^{14}$ sec & 
0.94 & 4.89 & 2.18 & 1.14 & 1.02 & 0.345 & 81.1 & 5.45 \\
& 
& & (1.08) & (0.913) & (0.972) & (1.01) & (1.55) & (1.03) \\
&&&&&&&&\\
& 
0.453 & 3.42 & 0.539 & 0.349 & 0.273 & 0.190 & 80.4 & 5.35 \\
&
& & (0.770) & (0.891) & (0.844) & (0.978) & (1.58) & (1.02) \\ 
\hline
$m_h = 10$ eV, $t_d = 10^{15}$ sec & 
1.12 & 10.1 & 2.25 & 1.40 & 1.11 & 0.242 & 45.6 & 7.55 \\
& 
& & (1.12) & (1.12) & (1.05) & (0.712) & (0.873) & (1.43) \\
&&&&&&&&\\
& 
0.825 & 8.17 & 0.727 & 0.739 & 0.400 & 0.173 & 43.8 & 6.74 \\
&
& & (1.04) & (1.89) & (1.24) & (0.889) & (0.863) & (1.29) \\ 
\hline
$m_h = 10$ eV, $t_d = 10^{16}$ sec & 
3.25 & 13.2 & 2.88 & 2.05 & 1.50 & 0.363 & 28.4 & 3.90 \\
& 
& & (1.43) & (1.64) & (1.42) & (1.07) & (0.544) & (0.737) \\
&&&&&&&&\\
& 
2.07 & 8.27 & 0.973 & 1.22 & 0.483 & 0.237 & 21.4 & 3.62 \\
&
& & (1.39) & (3.11) & (1.49) & (1.21) & (0.421) & (0.691) \\ 
\hline
$m_h = 3.16$ eV, $t_d = 10^{14}$ sec &
2.23 & 16.6 & 2.08 & 1.14 & 0.997 & 0.570 & 84.2 & 6.14 \\
& 
& & (1.03) & (0.910) & (0.945) & (1.68) & (1.61) & (1.16) \\
&&&&&&&&\\
& 
1.21 & 12.0 & 0.634 & 0.355 & 0.280 & 0.338 & 73.59 & 5.89 \\
&
& & (0.906) & (0.907) & (0.866) & (1.74) & (1.45) & (1.12) \\ 
\hline
$m_h = 3.16$ eV, $t_d = 10^{15}$ sec & 
3.24 & 28.6 & 2.10 & 1.28 & 1.09 & 0.407 & 67.5 & 6.25 \\
& 
& & (1.05) & (1.02) & (1.03) & (1.20) & (1.29) & (1.18) \\
&&&&&&&&\\
& 
2.48 & 20.7 & 0.726 & 0.666 & 0.362 & 0.273 & 66.6 & 5.78 \\
&
& & (1.04) & (1.70) & (1.12) & (1.41) & (1.31) & (1.10) \\ 
\hline
$m_h = 3.16$ eV, $t_d = 10^{16}$ sec &
5.35 & 18.9 & 2.62 & 1.50 & 1.35 & 0.524 & 46.6 & 6.65 \\
& 
& & (1.31) & (1.19) & (1.28) & (1.54) & (0.891) & (1.26) \\
&&&&&&&&\\
& 
3.36 & 9.63 & 0.876 & 0.749 & 0.434 & 0.318 & 36.1 & 6.40 \\
&
& & (1.25) & (1.91) & (1.34) & (1.63) & (0.712) & (1.22) \\ 
\hline
\end{tabular}
\end{center}
\label{tb:stat_late_pol}
\caption{Same as last table, but for Planck.}
\end{table}

The results for several almost-stable scenarios are shown in
Tables~\ref{tb:stat_stableMap} and \ref{tb:stat_stablePlanck}. The relative
degradations in the non-neutrino parameters become worse as $y$ increases because
there the CMB anisotropy starts to be affected by the decay products. For very low
values of $y$, the uncertainty degradations are the same as the case for stable
neutrinos, with $m_h$ as the sole additional cosmic parameter. A prominent feature of
the data here is that for both MAP and Planck, $\delta y \gg y$ for $y \ll 1$; it is
impossible to use the CMB to probe neutrino decays in the almost-stable limit. This
is because the late-ISW feature is imprinted at very low values of $l$ where the
cosmic variance is high and where there are few $l$'s to measure.

\begin{table}
\begin{center}
\begin{tabular}{|c|c|c|c|c|c|c|c|c|} \hline
Model & $m_h$ & $y$ & $\Omega_B$ & $\Omega_\Lambda$ & $h$ & $n_s$ &
$\tau_\ast$ & $Q$ 
\\ \hline
$m_h = 1.0$ eV, $y = 0.1$ & 
23.9 & 1050.  & 9.29 & 7.79 & 5.00 & 1.60 & 113. & 7.23 \\
&
& & (1.81) & (1.94) & (1.41) & (1.09) & (1.42) & (1.26) \\
&&&&&&&&\\
&
12.7 & 898.  & 5.11 & 5.34 & 2.66 & 0.731 & 88.9 & 6.81 \\
&
& & (2.25) & (3.12) & (1.82) & (1.10) & (1.54) & (1.27) \\
\hline
$m_h = 1.0$ eV, $y=1.0$ &
25.5 & 349.3  & 14.9 & 11.6 & 7.43 & 1.72 & 149. & 7.29 \\
&
& & (2.89) & (2.90) & (2.09) & (1.17) & (1.86) & (1.27) \\
&&&&&&&&\\
&
13.4 & 203.  & 6.20 & 5.45 & 3.08 & 0.903 & 95.9 & 6.39 \\
&
& & (2.74) & (3.18) & (2.11) & (1.36) & (1.66) & (1.19) \\
\hline
$m_h = 3.16$ eV, $y=0.1$ &
148. & 394.  & 18.9 & 6.68 & 5.90 & 3.83 & 249. & 7.59 \\
&
& & (3.67) & (1.66) & (1.66) & (2.60) & (3.12) & (1.32) \\
&&&&&&&&\\
&
15.5 & 331.  & 4.61 & 5.69 & 2.46 & 0.837 & 99.2 & 7.47 \\
&
& & (2.03) & (3.32) & (1.68) & (1.26) & (1.71) & (1.39) \\
\hline
$m_h = 3.16$ eV, $y=1.0$ &
210. & 252.  & 31.5 & 11.0 & 11.3 & 5.73 & 316. & 6.81 \\
&
& & (6.10) & (2.75) & (3.17) & (3.91) & (3.94) & (1.18) \\
&&&&&&&&\\
&
17.1 & 67.5  & 6.40 & 6.20 & 3.27 & 1.09 & 110. & 6.09 \\
&
& & (2.82) & (3.62) & (2.23) & (1.63) & (1.91) & (1.14) \\
\hline
\end{tabular}
\end{center}
\label{tb:stat_stableMap}
\caption{Using the CMB to measure $m_h$ and $y$ for nearly stable neutrinos, for
MAP. The statistical uncertainties on the cosmic parameters, $\delta
\lambda_i / \lambda_i$, in percent, are shown for several models. The number in
parenthesis is the ratio of the uncertainty to the uncertainty for \lcdm. For each
model the top row of data is for temperature data only; the bottom row includes
polarization.}
\end{table}

\begin{table}
\begin{center}
\begin{tabular}{|c|c|c|c|c|c|c|c|c|} \hline
Model & $m_h$ & $y$ & $\Omega_B$ & $\Omega_\Lambda$ & $h$ & $n_s$ &
$\tau_\ast$ & $Q$ 
\\ \hline
$m_h = 1.0$ eV, $y = 0.1$ & 
8.16 & 291.  & 2.88 & 2.17 & 1.44 & 0.410 & 57.1 & 5.47 \\
&
& & (1.43) & (1.72) & (1.37) & (1.20) & (1.09) & (1.03) \\
&&&&&&&&\\
&
4.23 & 152.  & 0.950 & 0.909 & 0.465 & 0.227 & 52.5 & 5.25 \\
&
& & (1.36) & (2.32) & (1.44) & (1.17) & (1.04) & (1.02) \\
\hline
$m_h = 1.0$ eV, $y=1.0$ &
10.5 & 128.  & 3.51 & 3.21 & 1.82 & 0.540 & 69.7 & 5.79 \\
&
& & (1.75) & (2.56) & (1.72) & (1.59) & (1.33) & (1.09) \\
&&&&&&&&\\
&
5.05 & 56.4  & 1.15 & 1.24 & 0.575 & 0.289 & 51.9 & 5.39 \\
&
& & (1.64) & (3.16) & (1.78) & (1.49) & (1.02) & (1.03) \\
\hline
$m_h = 3.16$ eV, $y=0.1$ &
4.64 & 159.  & 3.22 & 3.15 & 1.69 & 0.550 & 66.4 & 5.73 \\
&
& & (1.60) & (2.51) & (1.60) & (1.62) & (1.27) & (1.08) \\
&&&&&&&&\\
&
2.94 & 62.8  & 1.12 & 1.19 & 0.576 & 0.228 & 49.6 & 5.27 \\
&
& & (1.60) & (3.04) & (1.78) & (1.17) & (0.978) & (1.01) \\
\hline
$m_h = 3.16$ eV, $y=1.0$ &
5.74 & 51.0  & 4.60 & 4.37 & 2.38 & 0.686 & 83.7 & 5.46 \\
&
& & (2.29) & (3.48) & (2.25) & (2.02) & (1.60) & (1.03) \\
&&&&&&&&\\
&
3.67 & 20.6  & 1.40 & 1.50 & 0.727 & 0.389 & 51.5 & 5.03 \\
&
& & (2.01) & (3.82) & (2.25) & (2.00) & (1.01) & (0.959) \\
\hline
\end{tabular}
\end{center}
\label{tb:stat_stablePlanck}
\caption{Same as Tab.~\ref{tb:stat_stableMap}, but for Planck.}
\end{table}

\section{Summary}

The goal of this work was a study of using anisotropy in the CMB to constrain the
physics of neutrinos that decay into non-interacting daughter products. We presented
the formalism required to compute the CMB anisotropy spectra in these models. This
required calculating the energy densities and the perturbations in the decaying
neutrino and its decay products, and incorporating this physics into the CMBFAST
code~\cite{Seljak96}. We divided the decaying neutrino parameter space into regions,
delineated by significant physical scales, and discussed the physics behind the CMB
spectra in each region. An enhanced early or late integrated-ISW effect is the main
effect for most of the neutrino parameter space.

We then developed analytic methods, valid in the linear regime, to determine when a
model is distinguishable from some canonical model like \lcdm. With
temperature data alone MAP can distinguish stable neutrino models from \lcdm if the
neutrino mass $m_h \gtrsim 2$ eV. Adding polarization data, $m_h \gtrsim 0.5$ eV is
distinguishable. Planck can distinguish $m_h \gtrsim 0.5$ eV with temperature alone,
and $m_h > 0.25$ eV with polarization. MAP without polarization can distinguish
out-of-equilibrium, early-decaying models as long as $(m_h/\rm{MeV})^2 \,
t_d/\rm{sec} \gtrsim 230$, and with polarization if $(m_h/\rm{MeV})^2 \, t_d/\rm{sec}
\gtrsim 150$. For Planck without polarization, models with $(m_h/\rm{MeV})^2 \,
t_d/\rm{sec} \gtrsim 9$ are distinguishable, and with polarization if
$(m_h/\rm{MeV})^2 \, t_d/\rm{sec} \gtrsim 6$. Models in which neutrinos decay in
equilibrium are indistinguishable from \lcdm. Late-decaying models ($10^{13} \rm{sec}
\lesssim t_d \lesssim 4 \times 10^{17} \rm{sec}$) are distinguishable from \lcdm if $m_h
\gtrsim 5$ eV for MAP and $m_h \gtrsim 2$ eV for Planck. 

Next, we studied the use of future CMB satellite data to measure cosmic parameters,
including neutrino properties. The sensitivity to neutrino parameters depends
strongly on the parameters themselves. We found that including neutrino parameters in
a model significantly degrades the sensitivity to $\Omega_B$, $h$, and $n_s$, and
that the degradation is worse for MAP than Planck. For models whose CMB spectra are
not close to \lcdm, the situation is less simple, but the sensitivities to cosmic
parameters are usually less than for the canonical case. For early-decaying models,
the sensitivities to most non-neutrino parameters decreases as $\alpha$ increases. In
addition, we calculated the set of models (for early-decaying neutrinos, for now),
where the statistical uncertainty in the neutrino parameters is low enough relative
to the parameters themselves, to count as a detection of decaying neutrinos. For
early-decaying neutrinos, MAP with can achieve this if $\alpha
\gtrsim 10$ with temperature information alone, and if $\alpha
\gtrsim 3$ with polarization data. The equivalent sensitivities for Planck are for
$\alpha \gtrsim 1$ with temperature information alone, and $\alpha \gtrsim 0.8$ with
polarization data

Although presented in the context of decaying neutrino cosmologies, the techniques
developed here could easily be extended to more generic scenarios involving decaying
particles which decay into sterile daughter products. The main difference in the
calculation would be in determining the particle's initial abundance (the
relativistic decoupling of the decaying neutrino simplifies the calculation in this
case). Given this, the equations for the evolution of the densities and perturbations
would be the same as for decaying neutrinos.

In conclusion, future CMB observations promise to provide a powerful probe of
neutrino physics, over a wide range of parameter space not easily accessed by other
means. A couple of caveats are in order. First, this investigation was preliminary in
nature. Cosmic variance limited data is a best case scenario; real-world issues like
foreground subtraction will complicate the actual data analysis. Hopefully, the data
from MAP and Planck will approach this ideal. Second, the real world CMB anisotropy
might look nothing like any variant of CDM, with or without decaying neutrinos. In
this case, of course, the analysis presented here would no longer be valid; one would
first have to understand the background cosmology before going on to study the impact
of decaying neutrinos.

\acknowledgements
Thanks to Michael Turner for reviewing this paper and providing many helpful
comments. Thanks to Scott Dodelson, Robert Scherrer, Manoj Kaplinghat and Lloyd Knox
for helpful discussions. Thanks to Uros Seljak and Matthias Zaldarriaga for the use
of the CMBFAST code. This work was supported through the DOE at Chicago and Fermilab
and by NASA grant NAG 5-7092 (at Fermilab).

\appendix

\section*{Distinguishability of Models} 
\label{ap:dist} 

This appendix deals with the question of how to determine whether or not the CMB
spectra from massive decaying neutrino models are distinguishable from some baseline
model like \lcdm, which does not contain decaying neutrinos. Here we will restrict
ourselves to the case where the decaying neutrino model produces CMB spectra that
are only slightly different from the baseline. 

We consider the following scenario. The universe actually contains decaying
neutrinos, but the experimental data is analyzed without considering this
possibility: the set of \cps does not include $m_h$ or $t_d$. As a result, two things
can happen. One, the \cps measured will in general be unequal to the true \cps, i.e.,
the results will be biased. Two, the best-fit spectra may be a poor fit. If, for
example, the presence of the decaying neutrinos changed the spectrum in exactly the
same way as adding a little extra baryon density, then the measured baryon density
would be biased, but the best fit model would fit very well. It would be impossible
to disentangle the decaying neutrino signature from the data. We will call a model
{\em distinguishable} if the best fit model is a poor one.

To be more quantitative, we need to work through how one measures the \cps from the
data. Start with some definitions: let $\{\lambda_i\}$ be the set of \cps
considered. Here, $i=1\ldots N$, with $N$ the total number \cps. As mentioned, this
set does not include $m_h$ or $t_d$. Let $\{\tilde{\lambda}_i\}$ be the set of true
\cps, and $\{\lambda_i^\prime\}$ be measured \cps. Finally, let $\{\delta
\lambda_i\}$ be the parameter biases induced by the decaying neutrino's, i.e.,
$\lambda_i^\prime \equiv \tilde{\lambda}_i + \delta
\lambda_i$. The measured \cps are determined by minimizing a $\chi^2$ statistic that
is a function of $\{\lambda_i\}$, given by Eqn.~\ref{eq:chi2}.  Here we will assume
that the experimental uncertainties are just cosmic variance up to some maximum value
of $l=l_{max}$, so that the covariance matrix is that given in Eqns.~\ref{eq:covdiag}
and \ref{eq:covoffdiag}.

We know that the solution for the measured \cps with no decaying neutrinos and no
noise is $\{\lambda_i\} = \{\tilde{\lambda}_i\}$. Now assume that the parameter
biases, $\delta \lambda_i$, are small enough so that the following holds:
\begin{equation}
C_{Xl}^{theory}(\{\lambda_i^\prime\}) \simeq 
C_{Xl}^{theory}(\{\tilde{\lambda}_i\}) + 
\frac{\partial C_{Xl}^{theory}}{\partial \lambda_k} \delta \lambda_k \,.
\label{eq:smalldc}
\end{equation}
If the experiment measures the temperature anisotropy only, then $X=T$. With
polarization information, $X = T,P,C$. This equation quantifies the statement that
the anisotropy is only slightly different than for \lcdm. If this holds we can solve
for the parameter biases:
\begin{equation}
\delta \lambda_k = \left(\alpha_{jk}\right)^{-1} 
  \sum_l \sum_{XY} 
  \frac{\partial C_{Xl}^{theory}}{\partial \lambda_j} \,
  V^{-1}_{XYl} \,
  S_{Yl}
\,, 
\label{eq:bias}
\end{equation}
where $S_{Xl}$ is the ``signal'':
\begin{equation}
S_{Xl} = C_{Xl}^{data} - C_{Xl}^{theory} \,,
\end{equation}
and $\alpha_{jk}$ is the Fisher matrix, given in Eqn.~\ref{eq:fm}. Finally, the
best-fit $\chi^2$ is given by
\begin{equation}
\chi^2_{min} = \sum_l \sum_{XY}
  \left(S_{Xl} - 
  \delta \lambda_j \frac{\partial C_{Xl}^{theory}}{\partial \lambda_j}
  \right)
  V^{-1}_{XYl}
  \left(S_{Yl} - 
  \delta \lambda_j \frac{\partial C_{Yl}^{theory}}{\partial \lambda_j}
  \right)
\end{equation}

To determine whether or not a model is distinguishable, we will need to understand
the statistical properties of $\left<\chi^2_{min}\right>$, considered as an ensemble
over different realizations of cosmic variance ``noise''\footnote{Of course cosmic
variance is not noise in the proper sense. It is an intrinsic part of the anisotropy,
and is separated this way for convenience.}. To develop the formalism for estimating
the contributions to $\chi^2_{min}$ from noise and signal, we break the data spectrum
$C^{data}_l$ into two components:
\begin{equation}
C^{data}_{Xl} = C^{data,0}_{Xl} + N_{Xl} \,,
\end{equation}
where $C^{data,0}_{Xl}$ is the decaying neutrino spectrum without noise, and $N_{Xl}$
is the noise. In a perfect experiment, $N_{Xl}$ is cosmic variance. The signal
$S_{Xl}$ breaks up similarly: $S_{Xl} = S^0_{Xl} + N_{Xl}$. Then the best-fit
$\chi^2$ can be written 
\begin{equation}
\chi^2_{min} = \sum_l \sum_{XY}
  \left(S_{Xl} - 
  \delta \lambda_j \frac{\partial C_{Xl}^{theory}}{\partial \lambda_j}
  + N_{Xl}
  \right)
  V^{-1}_{XYl}
  \left(S_{Yl} - 
  \delta \lambda_j \frac{\partial C_{Yl}^{theory}}{\partial \lambda_j}
  + N_{Yl}
  \right)
\,,
\label{eq:chi2bf}
\end{equation}
where $\delta \lambda_j$ is the CP bias for {\em noiseless} data, i.e., without
cosmic variance.

Consider an ensemble of experiments for a given decaying neutrino model. Each
experiment will have the same noiseless signal, but the noise, a random variable,
will be different each time. Therefore, the value of $\chi^2_{min}$ will also
vary. Associated with each value of $\chi^2_{min}$ is some probability that the \lcdm
model is allowed, denoted $a$.  This probability is just the 1 minus the cumulative
distribution function, $\mathcal{E}$, for the $\chi^2$ distribution with $l_{max}-1$
degrees of freedom, evaluated at $\chi^2_{min}$:
\begin{equation}
a(\chi^2_{min}) = 1 - {\mathcal E}(l_{max}-1,\chi^2_{min}) \,.
\end{equation}
Then the confidence level for \lcdm, denoted $\mathcal{C}$, can be expressed as a
convolution of $a$ with the probability $P(\chi^2_{min})$ of obtaining different
values of $\chi^2_{min}$,
\begin{equation}
{\mathcal C} = 1 - \int_0^\infty \, d\chi^2_{min} P(\chi^2_{min}) 
  \left[ 1-{\mathcal E}(l_{max}-1,\chi^2_{min}) \right] \,.
\label{eq:A}
\end{equation}
To proceed further, we need to understand the shape of $P(\chi^2_{min})$, which is
determined by the distribution of $N_l$.

We will treat the $N_l$'s as Gaussian random variables with zero mean and variance
determined by cosmic variance. In this limit $P(\chi^2_{min})$ is Gaussian
too. However, the $N_l$'s are not really Gaussianly distributed. A more realistic
treatment~\cite{Knox98} reveals that their distribution is closer to log-normal, with
large high-$N_l$ tails. The disagreement is greater for low values of $l$; for high
values, say with $l \gtrsim 50$, the distribution is approximately normal. There are
a couple of reasons why it is acceptable to approximate their distributions as
normal. First, most of the statistical weight in distinguishing models comes from
high values of $l$, because cosmic variance is smaller there and for experiments we
will be considering, with $l_{max} \sim 1000$, there are just more values of $l$ that
are large than small. Second, many different distributions, one for each $N_l$,
collectively determine the statistical properties of the $\chi^2_{min}$ distribution,
and as the number of contributions becomes large, $P(\chi^2_{min})$ will tend towards
a Gaussian. In this case $P(\chi^2_{min})$ is specified completely by its mean (or
expectation value) $\left<\chi^2_{min}\right>$, and its and variance
\begin{equation}
\sigma^2_\chi = \left< \left(\chi^2_{min} \right)^2 \right> -
\left<\chi^2_{min}\right>^2 \,.
\end{equation}

First, the mean. Expanding the quadratic in Eqn.~\ref{eq:chi2bf}, we will have terms
proportional to $N_{Xl}$ and $N_{Xl} N_{Yl}$ and terms independent of $N_{Xl}$. The
expectation of the linear term is zero, since $\left<N_{Xl}\right> = 0$, as $N_{Xl}$
is a Gaussian random variable with mean 0. For the quadratic term, $\left<N_{Xl}
N_{Yl} \right> = 2 C^2_{Xl} / (2l+1) \delta_{XY}$, where $\delta_{XY}$ is the
discrete delta function. Therefore, we find for the expectation value of
$\chi^2_{min}$,
\begin{equation}
\left<\chi^2_{min}\right> =  
  \sum_l \sum_{XY} 
  \left(S^0_{Xl} - 
    \delta \lambda_j \frac{\partial C_{Xl}^{theory}}{\partial \lambda_j}
  \right)
  V^{-1}_{XYl}
  \left(S^0_{Yl} - 
    \delta \lambda_j \frac{\partial C_{Yl}^{theory}}{\partial \lambda_j}
  \right) +
  \sum_l \sum_X V^{-1}_{XX} \frac{2 C_{Xl}^{theory}}{2l+1}
\,.
\end{equation}

This expression simplifies in certain cases. Namely, if temperature and polarization
data can be considered uncorrelated then $X = T,P$ and $V^{-1}_{XY}$ is a diagonal
matrix, with $V^{-1}_{XX} = (2l+1)/2C_{Xl}^2$. Then the second term on the right hand
side of the last equation is just equal to the number of terms in the sum,
$2(l_{max}-1)$, and the expectation value becomes
\begin{equation}
\left<\chi^2_{min}\right> =  2(l_{max}-1) +
  \sum_l \sum_{X=T,P} 
  \frac{2(l+1)}{2 C_{Xl}^2}
  \left(
    \frac{\partial C_{Yl}^{theory}}{\partial \lambda_j} - S^0_{Xl} 
  \right)^2
\end{equation}

The variance is an unholy mess. The second term on the right hand side is the square
of the mean. The first term looks like the following:
\begin{equation}
\left< \left(\chi^2_{min} \right)^2 \right> =
\left< 
  \sum_{lm} \sum_{WXYZ} 
  \left(N_{Xl}+D_{Xl}\right)
  V^{-1}_{XYl}
  \left(N_{Yl}+D_{Yl}\right)
  \left(N_{Wm}+D_{Wm}\right)
  V^{-1}_{WZm}
  \left(N_{Zm}+D_{Zm}\right)
\right> 
\end{equation}
where
\begin{equation}
D_{Xl} = 
S_{Xl}^0 - \delta \lambda_j \frac{\partial C^{theory}_{Xl}}{\partial \lambda_j} 
\end{equation}
doesn't depend on $N_{Xl}$. 

If the temperature and polarization are uncorrelated, this equation simplifies
considerably. The sum inside the brackets will contain different powers of $N_{Xl}$
and $N_{Xm}$, the objects whose expectation values are non-trivial. Note that if the
power of either $N_l$ or $N_m$ is odd, then that term's expectation value will
vanish. In addition, terms that involve only $N_l^2$ or $N_m^2$ have already been
discussed. This allows the expression to be greatly simplified:
\begin{eqnarray}
\left< \left(\chi^2_{min} \right)^2 \right> &=&
  \left( \sum_l \sum_{X=T,P} 
  \frac{2l+1}{2 C_{Xl}^2} D_{lM}^2 \right)^2 + 
  4 \left(l_{max}-1\right) 
  \left( \sum_l \sum_{X=T,P} 
  \frac{2l+1}{2 C_{Xl}^2} D_{lM}^2 \right) + 
\nonumber \\
&&
  \sum_{l,m} \sum_{X,Y=T,P}
  \frac{2l+1}{2 C_{Xl}^2}
  \frac{2m+1}{2 C_{Yl}^2}
  {\left< N_l^2\,N_m^2 \right>} \,.
\end{eqnarray}
The last term on the right hand side can be evaluated by noting the following
identities: $\left<N_{Xl}^2 N_{Xl}^2 \right> = 3 (2C_{Xl}^2/(2l+1))^2$, and
$\left<N_{Xl}^2 N_{Ym}^2 \right> = (2C_{Xl}^2/(2l+1))(2C_{Ym}^2/(2l+1))$ if $X\neq Y$
or $l \neq m$. Using these identities, we find a simple formula for the variance:
\begin{equation}
\sigma_\chi^2 = 4\left(l_{max}-1\right) \,.
\label{eq:sigma}
\end{equation}
Note that the formula depends only on the number of degrees of freedom and not on
$\left<\chi^2_{min}\right>$. Finally, we can express Eqn.~\ref{eq:A} in terms of the
probability distribution for $\chi^2_{min}$,
\begin{equation}
{\mathcal C} = 1 - \int_0^\infty \, d\chi^2
  \frac{ 1-{\mathcal E}(l_{max}-1,\chi^2)}{\sqrt{2\pi}\,\sigma_\chi} \,
  \exp\left[\frac{-\left(\chi^2-\left<\chi^2\right>\right)^2}{2\sigma^2_\chi}\right]
\,.
\label{eq:Afinal}
\end{equation}

\newpage
\bibliography{thesis}

\begin{thebibliography}{10}
\providecommand*{\bibinfo}[2]{#2}
\providecommand*{\eprint}[1]{#1}
\providecommand*{\url}[1]{#1}
\bibitem{Bunn97}
\bibinfo{author}{J.~R. Bond} and \bibinfo{author}{M.~White},
  \bibinfo{journal}{\apj} \bibinfo{volume}{\textbf{480}}, \bibinfo{pages}{6}
  (\bibinfo{date}{1997}).
\bibitem{Bernardis97}
\bibinfo{author}{P.~de~Bernardis} \emph{et~al.}, \bibinfo{journal}{\apj}
  \bibinfo{volume}{\textbf{480}}, \bibinfo{pages}{1} (\bibinfo{date}{1997}).
\bibitem{Lineweaver98}
\bibinfo{author}{C.~H. Lineweaver}, \bibinfo{journal}{\apj}
  \bibinfo{volume}{\textbf{505}}, \bibinfo{pages}{L69} (\bibinfo{date}{1998}).
\bibitem{Hancock98}
\bibinfo{author}{S.~Hancock} \emph{et~al.}, \bibinfo{journal}{\mnras}
  \bibinfo{volume}{\textbf{294}}, \bibinfo{pages}{1} (\bibinfo{date}{1998}).
\bibitem{Lesgourgues98}
\bibinfo{author}{J.~Lesgourgues} \emph{et~al.}, \bibinfo{journal}{\ph{9807019}}
   (\bibinfo{date}{1998}).
\bibitem{Bartlett98}
\bibinfo{author}{J.~Bartlett} \emph{et~al.}, \bibinfo{journal}{\ph{9804158}}
  (\bibinfo{date}{1998}).
\bibitem{Bond98}
\bibinfo{author}{J.~R. Bond} and \bibinfo{author}{A.~H. Jaffee},
  \bibinfo{journal}{\ph{9808043}}  (\bibinfo{date}{1998}).
\bibitem{Webster98}
\bibinfo{author}{A.~M. Webster}, \bibinfo{journal}{\apj}
  \bibinfo{volume}{\textbf{509}}, \bibinfo{pages}{L65} (\bibinfo{date}{1998}).
\bibitem{White98}
\bibinfo{author}{M.~White}, \bibinfo{journal}{\apj}
  \bibinfo{volume}{\textbf{506}}, \bibinfo{pages}{485} (\bibinfo{date}{1998}).
\bibitem{Ratra99}
\bibinfo{author}{B.~Ratra} \emph{et~al.}, \bibinfo{journal}{\apj}
  \bibinfo{volume}{\textbf{517}} (\bibinfo{date}{1999}).
\bibitem{Eisenstein98}
\bibinfo{author}{D.~Eisenstein}, \bibinfo{author}{W.~Hu}, and
  \bibinfo{author}{M.~Tegmark}, \bibinfo{journal}{\ph{9807130}}
  (\bibinfo{date}{1998}).
\bibitem{Tegmark99}
\bibinfo{author}{M.~Tegmark}, \bibinfo{journal}{\apj}
  \bibinfo{volume}{\textbf{514}} (\bibinfo{date}{1999}).
\bibitem{map}
\bibinfo{title}{\emph{http://map.gsfc.nasa.gov}}.
\bibitem{planck}
\bibinfo{title}{\emph{http://astro.estec.esa.nl/sa-general/projects/cobras/cob%
ras.html}}.
\bibitem{Lopez98a}
\bibinfo{author}{R.~E. Lopez}, \bibinfo{author}{S.~Dodelson},
  \bibinfo{author}{A.~Heckler}, and \bibinfo{author}{M.~S. Turner},
  \bibinfo{journal}{\prl} \bibinfo{volume}{\textbf{82}}, \bibinfo{pages}{3952}
  (\bibinfo{date}{1999}).
\bibitem{Kaplinghat98}
\bibinfo{author}{M.~Kaplinghat}, \bibinfo{author}{R.~J. Scherrer}, and
  \bibinfo{author}{M.~S. Turner}, \bibinfo{journal}{\prd}
  \bibinfo{volume}{\textbf{60}}, \bibinfo{pages}{023516}
  (\bibinfo{date}{1999}).
\bibitem{Kinney99}
\bibinfo{author}{W.~H. Kinney} and \bibinfo{author}{A.~Riotto},
  \bibinfo{journal}{\ph{9903459}}  (\bibinfo{date}{1999}).
\bibitem{Liddle98}
\bibinfo{author}{A.~Liddle}, \bibinfo{author}{A.~Mazumdar}, and
  \bibinfo{author}{J.~Barrow}, \bibinfo{journal}{\prd}
  \bibinfo{volume}{\textbf{58}}, \bibinfo{pages}{027302}
  (\bibinfo{date}{1998}).
\bibitem{Kawasaki93}
\bibinfo{author}{M.~Kawasaki} and \bibinfo{author}{H.-S. Kang},
  \bibinfo{journal}{Nuc. Phys. B} \bibinfo{volume}{\textbf{403}},
  \bibinfo{pages}{671} (\bibinfo{date}{1993}).
\bibitem{Dodelson94}
\bibinfo{author}{S.~Dodelson}, \bibinfo{author}{G.~Gyuk}, and
  \bibinfo{author}{M.~S. Turner}, \bibinfo{journal}{Phys. Rev. Lett.}
  \bibinfo{volume}{\textbf{72}}, \bibinfo{pages}{3754} (\bibinfo{date}{1994}).
\bibitem{Dodelson94a}
\bibinfo{author}{S.~Dodelson}, \bibinfo{author}{G.~Gyuk}, and
  \bibinfo{author}{M.~S. Turner}, \bibinfo{journal}{Phys. Rev. D}
  \bibinfo{volume}{\textbf{49}}, \bibinfo{pages}{5068} (\bibinfo{date}{1994}).
\bibitem{Madsen92}
\bibinfo{author}{J.~Madsen}, \bibinfo{journal}{Phys. Rev. Lett.}
  \bibinfo{volume}{\textbf{69}}, \bibinfo{pages}{571} (\bibinfo{date}{1992}).
\bibitem{Kawasaki94}
\bibinfo{author}{M.~Kawasaki} \emph{et~al.}, \bibinfo{journal}{Nuc. Phys. B}
  \bibinfo{volume}{\textbf{419}}, \bibinfo{pages}{105} (\bibinfo{date}{1994}).
\bibitem{Bond83}
\bibinfo{author}{J.~Bond} and \bibinfo{author}{A.~Szalay},
  \bibinfo{journal}{Astrophys. J.} \bibinfo{volume}{\textbf{274}},
  \bibinfo{pages}{443} (\bibinfo{date}{1983}).
\bibitem{Steigman85}
\bibinfo{author}{G.~Steigman} and \bibinfo{author}{M.~S. Turner},
  \bibinfo{journal}{Nuc. Phys. B} \bibinfo{volume}{\textbf{253}},
  \bibinfo{pages}{375} (\bibinfo{date}{1985}).
\bibitem{Bond91}
\bibinfo{author}{J.~Bond} and \bibinfo{author}{G.~Efstathiou},
  \bibinfo{journal}{Phys. Lett. B} \bibinfo{volume}{\textbf{265}},
  \bibinfo{pages}{245} (\bibinfo{date}{1991}).
\bibitem{White95}
\bibinfo{author}{M.~White}, \bibinfo{author}{G.~Gelmini}, and
  \bibinfo{author}{J.~Silk}, \bibinfo{journal}{Phys. Rev. D}
  \bibinfo{volume}{\textbf{51}}, \bibinfo{pages}{2669} (\bibinfo{date}{1995}).
\bibitem{Bharadwaj98}
\bibinfo{author}{S.~Bharadwaj} and \bibinfo{author}{S.~K. Sethi},
  \bibinfo{journal}{Astrophys. J. Suppl.} \bibinfo{volume}{\textbf{114}},
  \bibinfo{pages}{37} (\bibinfo{date}{1998}).
\bibitem{Hannestad98}
\bibinfo{author}{S.~Hannestad}, \bibinfo{journal}{Phys. Lett. B}
  \bibinfo{volume}{\textbf{431}}, \bibinfo{pages}{363} (\bibinfo{date}{1998}).
\bibitem{Hannestad98b}
\bibinfo{author}{S.~Hannestad}, \bibinfo{journal}{\prl}
  \bibinfo{volume}{\textbf{80}}, \bibinfo{pages}{4621} (\bibinfo{date}{1998}).
\bibitem{Hannestad99}
\bibinfo{author}{S.~Hannestad}, \bibinfo{journal}{\prd}
  \bibinfo{volume}{\textbf{59}}, \bibinfo{pages}{105020}
  (\bibinfo{date}{1999}).
\bibitem{Lopez99a}
\bibinfo{author}{M.~Kaplinghat}, \bibinfo{author}{R.~E. Lopez},
  \bibinfo{author}{S.~Dodelson}, and \bibinfo{author}{R.~J. Scherrer},
  \bibinfo{journal}{\ph{9907388}}  (\bibinfo{date}{1999}).
\bibitem{Lopez98}
\bibinfo{author}{R.~E. Lopez}, \bibinfo{author}{S.~Dodelson},
  \bibinfo{author}{R.~J. Scherrer}, and \bibinfo{author}{M.~S. Turner},
  \bibinfo{journal}{\prl} \bibinfo{volume}{\textbf{81}}, \bibinfo{pages}{3075}
  (\bibinfo{date}{1998}).
\bibitem{Raffelt99}
\bibinfo{author}{G.~G. Raffelt}, \bibinfo{journal}{hep-ph/9902271}
  (\bibinfo{date}{1999}).
\bibitem{Raffelt96}
\bibinfo{author}{G.~G. Raffelt}, \bibinfo{title}{\emph{Stars as Laboratories
  for Fundamental Physics}} (\bibinfo{publisher}{University of Chicago Press},
  \bibinfo{year}{1996}).
\bibitem{Mohapatra98}
\bibinfo{author}{R.~N. Mohapatra} and \bibinfo{author}{P.~B. Pal},
  \bibinfo{title}{\emph{Massive Neutrinos in Physics and Astrophysics}}
  (\bibinfo{publisher}{World Scientific}, Singapore, \bibinfo{year}{1998}), 2nd
  ed.
\bibitem{Wilczek82}
\bibinfo{author}{F.~Wilczek}, \bibinfo{journal}{\prl}
  \bibinfo{volume}{\textbf{49}}, \bibinfo{pages}{1549} (\bibinfo{date}{1982}).
\bibitem{Reiss82}
\bibinfo{author}{D.~B. Reiss}, \bibinfo{journal}{Phys. Lett. B}
  \bibinfo{volume}{\textbf{115}}, \bibinfo{pages}{217} (\bibinfo{date}{1982}).
\bibitem{Gelmini83}
\bibinfo{author}{G.~B. Gelmini}, \bibinfo{author}{S.~Nussinov}, and
  \bibinfo{author}{T.~Yanagida}, \bibinfo{journal}{Nuc. Phys. B}
  \bibinfo{volume}{\textbf{219}}, \bibinfo{pages}{31} (\bibinfo{date}{1983}).
\bibitem{argus95}
\bibinfo{author}{H.~Albrecht} \emph{et~al.}, \bibinfo{journal}{Z. Phys. C}
  \bibinfo{volume}{\textbf{68}}, \bibinfo{pages}{25} (\bibinfo{date}{1995}).
\bibitem{Kamionkowski97}
\bibinfo{author}{M.~Kamionkowski}, \bibinfo{author}{A.~Kosowsky}, and
  \bibinfo{author}{A.~Stebbins}, \bibinfo{journal}{\prd}
  \bibinfo{volume}{\textbf{55}}, \bibinfo{pages}{7368} (\bibinfo{date}{1997}).
\bibitem{Zaldarriaga97a}
\bibinfo{author}{M.~Zaldarriaga} and \bibinfo{author}{U.~Seljak},
  \bibinfo{journal}{\prd} \bibinfo{volume}{\textbf{55}}, \bibinfo{pages}{1830}
  (\bibinfo{date}{1997}).
\bibitem{Zaldarriaga98}
\bibinfo{author}{M.~Zaldarriaga} and \bibinfo{author}{U.~Seljak},
  \bibinfo{journal}{\prd} \bibinfo{volume}{\textbf{58}},
  \bibinfo{pages}{023003} (\bibinfo{date}{1998}).
\bibitem{Lyth98}
\bibinfo{author}{D.~H. Lyth} and \bibinfo{author}{A.~Riotto},
  \bibinfo{journal}{Phys. Rept.} \bibinfo{volume}{\textbf{314}},
  \bibinfo{pages}{1} (\bibinfo{date}{1998}).
\bibitem{Ma95}
\bibinfo{author}{C.-P. Ma} and \bibinfo{author}{E.~Bertschinger},
  \bibinfo{journal}{\apj} \bibinfo{volume}{\textbf{455}}, \bibinfo{pages}{7}
  (\bibinfo{date}{1995}).
\bibitem{Kolb90}
\bibinfo{author}{E.~Kolb} and \bibinfo{author}{M.~Turner},
  \bibinfo{title}{\emph{The Early Universe}}
  (\bibinfo{publisher}{Addison-Wesley}, Reading, MA, \bibinfo{year}{1990}).
\bibitem{Seljak96}
\bibinfo{author}{U.~Seljak} and \bibinfo{author}{M.~Zaldarriaga},
  \bibinfo{journal}{Astrophys. J.} \bibinfo{volume}{\textbf{469}},
  \bibinfo{pages}{437} (\bibinfo{date}{1996}).
\bibitem{Jungman96}
\bibinfo{author}{G.~Jungman}, \bibinfo{author}{M.~Kamionkowski}, and
  \bibinfo{author}{A.~Kosowsky}, \bibinfo{journal}{\prd}
  \bibinfo{volume}{\textbf{54}}, \bibinfo{pages}{1332} (\bibinfo{date}{1996}).
\bibitem{Dodelson96}
\bibinfo{author}{S.~Dodelson}, \bibinfo{author}{E.~Gates}, and
  \bibinfo{author}{A.~Stebbins}, \bibinfo{journal}{Astrophys. J.}
  \bibinfo{volume}{\textbf{467}}, \bibinfo{pages}{10} (\bibinfo{date}{1996}).
\bibitem{Zaldarriaga97}
\bibinfo{author}{M.~Zaldarriaga}, \bibinfo{author}{D.~Spergel}, and
  \bibinfo{author}{U.~Seljak}, \bibinfo{journal}{\apj}
  \bibinfo{volume}{\textbf{488}}, \bibinfo{pages}{1} (\bibinfo{date}{1997}).
\bibitem{Bond97}
\bibinfo{author}{J.~R. Bond}, \bibinfo{author}{G.~Efstathiou}, and
  \bibinfo{author}{M.~Tegmark}, \bibinfo{journal}{\mnras}
  \bibinfo{volume}{\textbf{291}}, \bibinfo{pages}{L33} (\bibinfo{date}{1997}).
\bibitem{Tegmark97}
\bibinfo{author}{M.~Tegmark}, \bibinfo{author}{A.~N. Taylor}, and
  \bibinfo{author}{A.~F. Heavens}, \bibinfo{journal}{Astrophys. J.}
  \bibinfo{volume}{\textbf{480}}, \bibinfo{pages}{22} (\bibinfo{date}{1997}).
\bibitem{Press90}
\bibinfo{author}{W.~Press}, \bibinfo{author}{S.~Teukolsky},
  \bibinfo{author}{W.~Vetterling}, and \bibinfo{author}{B.~Flannery},
  \bibinfo{title}{\emph{Numerical Recipes in C}} (\bibinfo{publisher}{Cambridge
  University Press}, Cambridge, \bibinfo{year}{1990}).
\bibitem{Knox98}
\bibinfo{author}{J.~R. Bond}, \bibinfo{author}{A.~H. Jaffe}, and
  \bibinfo{author}{L.~E. Knox}, \bibinfo{journal}{\ph{9808264}}
  (\bibinfo{date}{1998}).

\end{thebibliography}

\end{document}